\documentclass[aps,pre,twocolumn,superscriptaddress,showpacs,floatfix,longbibliography]{revtex4-1}
\usepackage{dcolumn}
\usepackage{amsmath}
\usepackage{amssymb}
\usepackage{graphicx}
\usepackage{bm}
\usepackage[T1]{fontenc}
\usepackage{color}
\usepackage{url}
\usepackage[bf]{subfigure}
\usepackage{rotating}
\usepackage{scalefnt}
\usepackage{multirow}

\def\la{\left\langle\rule{0pt}{1em}}
\def\ra{\right\rangle}

\begin{document}

\title{Disease Localization in Multilayer Networks}

\author{Guilherme Ferraz de Arruda}
\affiliation{Departamento de Matem\'{a}tica Aplicada e Estat\'{i}stica, Instituto de Ci\^{e}ncias Matem\'{a}ticas e de Computa\c{c}\~{a}o,
Universidade de S\~{a}o Paulo - Campus de S\~{a}o Carlos, Caixa Postal 668,
13560-970 S\~{a}o Carlos, SP, Brazil.}
\affiliation{Institute for Biocomputation and Physics of Complex Systems (BIFI), University of Zaragoza, Zaragoza 50009, Spain}

\author{Emanuele Cozzo}
\affiliation{Institute for Biocomputation and Physics of Complex Systems (BIFI), University of Zaragoza, Zaragoza 50009, Spain}
\affiliation{Department of Theoretical Physics, University of Zaragoza, Zaragoza 50009, Spain}

\author{Tiago P. Peixoto}
\affiliation{Institut f\"ur Theoretische Physik, Universit\"at Bremen,
  Hochschulring 18, D-28359 Bremen, Germany}
\affiliation{ISI Foundation, Turin, Italy}

\author{Francisco A. Rodrigues}
\email{francisco@icmc.usp.br}
\affiliation{Departamento de Matem\'{a}tica Aplicada e Estat\'{i}stica, Instituto de Ci\^{e}ncias Matem\'{a}ticas e de Computa\c{c}\~{a}o,
Universidade de S\~{a}o Paulo - Campus de S\~{a}o Carlos, Caixa Postal 668,
13560-970 S\~{a}o Carlos, SP, Brazil.}

\author{Yamir Moreno}
\email{yamir.moreno@gmail.com}
\affiliation{Institute for Biocomputation and Physics of Complex Systems (BIFI), University of Zaragoza, Zaragoza 50009, Spain}
\affiliation{Department of Theoretical Physics, University of Zaragoza, Zaragoza 50009, Spain}
\affiliation{Complex Networks and Systems Lagrange Lab, Institute for Scientific Interchange, Turin, Italy}

\begin{abstract}
We present a continuous formulation of epidemic spreading on
multilayer networks using a tensorial representation, extending the models
of monoplex networks to this context. We derive analytical expressions
for the epidemic threshold of the SIS and SIR dynamics, as well as upper
and lower bounds for the disease prevalence in the steady state for the SIS scenario. Using the quasi-stationary state
 method we numerically show the existence of disease localization and the emergence of two or more susceptibility peaks, which are characterized analytically and numerically through the inverse participation ratio. Furthermore, when mapping the critical dynamics to an eigenvalue problem, we observe a characteristic
transition in the eigenvalue spectra of the supra-contact tensor as a
function of the ratio of two spreading rates: if the rate at which the disease spreads within a layer is comparable to the spreading rate across layers, the individual spectra of each layer merge with the coupling between layers. Finally, we verified the barrier effect, i.e., for three-layer configuration, when the layer with the largest eigenvalue is located at the center of the line, it can effectively act as a barrier to the disease. The formalism introduced here provides a unifying mathematical approach to disease contagion in multiplex systems opening new possibilities for the study of spreading processes. 
\end{abstract}

\maketitle

% \tableofcontents

\section{Introduction}

Epidemic like spreading processes are paradigmatic, as they can describe not
only the temporal unfolding and evolution of diseases, but also of ideas, information
and rumors in fields as diverse as biological, information and social sciences \cite{pastor-satorras_epidemic_2015}. Due to
their fundamental nature and simplicity, two particular models have received
special attention by the scientific community, the
susceptible-infected-susceptible (SIS) and the
susceptible-infected-recovered (SIR). In both models, an infected
individual spreads the disease to its neighbors at a given (spreading) rate and infected individuals recover at some other rate. The difference between both scenarios lies in the fact that in the SIS case, once recovered, infected individuals can catch the disease again, and, therefore, they go back to the susceptible state. On the contrary, in the SIR model, recovered individuals are supposed to acquire permanent immunity and do not play any active role in the spreading process anymore. There are many other variations of these two models, including more realistic and intricate compartmental models~\cite{pastor-satorras_epidemic_2015}. However, these two schemes are sufficient to capture the main phenomenology of disease dynamics --- and many other contagion like processes --- including the onset of epidemics, while remaining simple.

Originally, the modeling of diseases was confined to homogeneous systems, where any pair of individuals have the same contact probability~\cite{Anderson92, Barrat08:book}. However, most real-world networks are heterogeneously organized, which led to reexamine previous results considering non-trivial patterns among individuals, such as power-law degree distributions~\cite{Newman03, Boccaletti06:PR,Costa07:AP}. In~\cite{Satorras2001}, the authors presented the heterogeneous mean-field approach (HMF), showing that the epidemic threshold tends to zero in the thermodynamic limit on scale-free networks when they characteristic exponent is less than 3. This observation about the role of network organization changed completely our previous understanding of how disease outbreaks should be modeled and controlled, placing the focus of attention not only into new ways to model disease dynamics, but also into the incorporation of real contact patterns in the dynamical settings~\cite{Boguna2002, Barrat08:book, Newman010:book, Mieghem2012, Ferreira2012}.

Since then, many computational and theoretical frameworks have been proposed, which undoubtedly had made the modeling of disease contagion an active area of research and have provided new phenomenological insights and accurate methods for the study of real outbreaks. For instance, instead of the HMF approach, one can adopt the quenched mean field
(QMF) method, where a specific network is fixed and the dynamics is
modeled in terms of nodal probabilities~\cite{Wang03,
Mieghem09}. The results obtained with the latter approach show that the
epidemic threshold depends on the inverse of the leading eigenvalue of
the adjacency matrix~\cite{Wang03, Mieghem09} --- a similar result was also
obtained using a discrete Markov chain approach~\cite{Gomez2010}.  Other scenarios explored recently include the case of temporal networks~\cite{Holme2012, Colizza2015}, competing and interacting diseases~\cite{Newman2005, Pedersen2007, Rohani2008, Louis2012, Vespignani2013, Caterina2014, Yamir2014} as well as the inclusion of human behavioral responses~\cite{Funk2009, Funk2010, Vespignani2011}.

However, the vast majority of the works so far deal with single-layered networks, despite the fact that many real systems exhibit a large degree of interconnectivity and hence should be modeled as multilayer networks~\cite{Kivela2014}. Such systems represent multimodal, multicategorical or temporal interactions, as for instance social relations, the ecosystem formed by different online social networks or modern transportation systems \cite{Kivela2014}. Cozzo et al.~\cite{Cozzo2013} showed that disregarding the multilayer structure can lead to misleading conclusions, missing fundamental aspects of the critical dynamics of spreading-like processes. Such findings reinforce the importance of a more detailed
investigation of contagion processes on multilayer networks. Here, we develop a theoretical and computational framework for the analysis of disease spreading, generalizing the results of Ref.~\cite{Mieghem09} to multilayer networks. A continuous counterpart to the model presented in \cite{Cozzo2013} is provided in terms of the tensorial notation introduced in \cite{DeDomenico2013}. Our methodology allows for several new results. First, we are able to write down in a compact form the equations describing the disease dynamics in a multilayer system. Secondly, we derive the corresponding epidemic thresholds for the SIS and SIR cases as well as establish bounds for the prevalence of the disease in the SIS scenario. We also identify previously unnoticed multiple susceptibility peaks and disease localization, which are traced back to the very topological nature of the system and described in terms of the eigenvalue spectra of the supra-contact tensor and the localization of eigenstates. 

The rest of the paper is organized as follows: we first formally define the concept of multilayer network, introducing the tensorial notation. Next we derive the equations describing the dynamics of the disease for the SIS scheme, calculating the upper and lower bounds for the prevalence of the disease in the steady state, followed by the analytical expression for the epidemic threshold, which is also derived for the SIR model. Furthermore, we use the results in~\cite{Garcia2014} to define some constraints on the critical point. In addition, we explore the notion of localization of eigenstates, formerly applied on epidemic spreading in~\cite{Goltsev2012}, to inspect localization transitions, which were verified by multiple susceptibility peaks. Finally, we also present results from extensive numerical simulations considering multiplex networks with scale-free and scale-rich structures, computing their respective epidemic thresholds. Finally, we present our conclusions in the last section.

\section{Continuous formulation for multilayer epidemic spreading} \label{sec:teo}

Multilayer networks have been shown to better describe interdependent systems. Mathematically, they can be described by either generalizing the matrix representation and formalism \cite{Kivela2014} or by encoding the system's topology in a tensorial representation, which was recently proposed~\cite{DeDomenico2013} and first applied to describe a dynamical process in~\cite{deArruda2016}. Here, we use the latter framework to formulate a continuous time Markov chain model that describes the evolution of an epidemic processes.

\subsection{Tensorial representation}

Tensors are elegant mathematical objects that generalize the concepts of scalars, vectors and matrices. A tensorial representation provides a natural and concise framework for modeling and solving multidimensional problems and is widely used in different fields, from linear algebra to physics. In particular, general relativity is completely formulated under the tensorial notation. Here we use the representation formerly presented
in~\cite{DeDomenico2013}. We also adopt the Einstein summation convention,
in order to have more compact equations: if two indices are repeated,
where one is a superscript and the other a subscript, then such operation
implies a summation. Aside from that, the result is a tensor whose rank
lowers by 2. For instance, $A^{\alpha}_{\beta} A_{\alpha}^{\gamma} =
\sum_\alpha A^{\alpha}_{\beta} A_{\alpha}^{\gamma}$. In our notation we
use greek letters to indicate the components of a tensor. In addition,
we use tilde ($\tilde{\cdotp}$) to denote the components related to the
layers, with dimension $m$, while the components without tilde have
dimension $n$ and are related to the nodes.

A multilayer network is represented as the fourth-order adjacency tensor
$M \in \mathbb{R}^{n \times n \times m \times m}$, which can represent several relations between nodes~\cite{DeDomenico2013}
\begin{equation}
\begin{split}
 M_{\beta \tilde{\gamma}}^{\alpha \tilde{\delta}} &= \sum_{\tilde{h},\tilde{k} = 1}^m C^{\alpha}_{\beta}(\tilde{h}\tilde{k}) E^{\tilde{\delta}}_{\tilde{\gamma}}(\tilde{h}\tilde{k}) = \\
  &= \sum_{\tilde{h},\tilde{k} = 1}^m \sum_{i,j = 1}^n w_{ij} (\tilde{h}\tilde{k}) \mathcal{E}_{\beta \tilde{\gamma}}^{\alpha \tilde{\delta}}(ij\tilde{h}\tilde{k}),
\end{split}
\end{equation}
where $E_{\tilde{\delta}}^{\tilde{\gamma}}(\tilde{h}\tilde{k}) \in
\mathbb{R}^{m \times m}$ and $\mathcal{E}_{\beta \tilde{\gamma}}^{\alpha \tilde{\delta}}(ij\tilde{h}\tilde{k}) \in
\mathbb{R}^{n \times n \times m \times m}$ indicate the tensor in its
respective canonical basis. Observe that we can extract one layer by projecting the tensor $M_{\beta \tilde{\gamma}}^{\alpha \tilde{\delta}}$
to the canonical tensor
$E_{\tilde{\delta}}^{\tilde{\gamma}}(\tilde{r}\tilde{r})$. Formally,
from~\cite{DeDomenico2013} we have
\begin{equation}
 M_{\beta \tilde{\gamma}}^{\alpha \tilde{\delta}} E_{\tilde{\delta}}^{\tilde{\gamma}}(\tilde{r}\tilde{r}) = C^{\alpha}_{\beta}(\tilde{r}\tilde{r}) = A^{\alpha}_{\beta}(\tilde{r}),
\end{equation}
where $\tilde{r} \in \{ 1, 2, ..., m \}$ is the selected layer and $A^{\alpha}_{\beta}(\tilde{r})$ is the adjacency matrix (rank-2 tensor). Moreover, aiming at having more compact and clear equations we define the all-one tensors $u_\alpha \in \mathbb{R}^n$ and $U^{\beta \tilde{\delta}} \in \mathbb{R}^{n \times m}$. Here, we restrict our analysis to multilayer networks with a diagonal coupling~\cite{Kivela2014}. In other words, each node can have at most one counterpart on the other layers. In addition, for simplicity, we focus on unweighted and undirected connected networks, in which there is a path from each node to all other nodes. For complementary information about the tensorial representation, its projections and the generalization of the eigenvalue problem, see Appendix~\ref{sec:app_tensor}.

\subsection{The Susceptible-Infected-Susceptible (SIS) model} \label{sec:sis}

Despite its simplicity, the susceptible-infected-susceptible (SIS) and susceptible-infected-recovered (SIR) models capture the main features of disease spreading~\cite{pastor-satorras_epidemic_2015}. In this section we focus on the first order approximation of the SIS model. Additionally, we present some aspects of the SIS exact formulation on Appendix~\ref{sec:app_sis_comp} and a brief analysis of the SIR model on Appendix~\ref{sec:app_sir}. 

We model the SIS disease dynamics associating a Poisson process to each of the elementary dynamical transitions: intra and inter layer spreading and the recovery from the infected state. The first two processes are associated to the edges of the graph and are characterized by the parameters $\lambda$ and $\eta$, respectively. The latter transition is modeled in the node, also via a Poisson process with parameter $\delta$. Using the tensorial notation defined above, the equations describing the systems dynamics read as
\begin{equation} \label{eq:SIS_tensor}
\dfrac{d X_{\beta \tilde{\delta}}}{dt} = - \mu X_{\beta \tilde{\delta}} +\left( 1 - X_{\beta \tilde{\delta}} \right) \lambda \mathcal{R}_{\beta \tilde{\delta}}^{\alpha \tilde{\gamma}}(\lambda, \eta)  X_{\alpha \tilde{\gamma}},
\end{equation}
where the supra contact tensor is defined as
\begin{equation} \label{eq:adj_tensor}
\mathcal{R}_{\beta \tilde{\delta}}^{\alpha \tilde{\gamma}}(\lambda, \eta) = M_{\beta \tilde{\sigma}}^{\alpha \tilde{\eta}} E^{\tilde{\sigma}}_{\tilde{\eta}}(\tilde{\gamma} \tilde{\delta})  \delta^{\tilde{\gamma}}_{\tilde{\delta}} +
\frac{\eta}{\lambda} M_{\beta \tilde{\sigma}}^{\alpha \tilde{\eta}} E^{\tilde{\sigma}}_{\tilde{\eta}}(\tilde{\gamma} \tilde{\delta}) (U^{\tilde{\gamma}}_{\tilde{\delta}} - \delta^{\tilde{\gamma}}_{\tilde{\delta}}),
\end{equation}
which encodes the contacts. It has a similar role as the matrix $R$ in~\cite{Cozzo2013}. Notice that we have implicitly assumed
that the random variables $X_{\beta \tilde{\delta}}$ are independent. Formally, if the state variable (Bernoulli random variable) $\boldsymbol{S_{\beta \tilde{\delta}}}$ is such that $\boldsymbol{S_{\beta \tilde{\delta}}} = 1$ when the node $\beta$ on layer $\tilde{\delta}$ is a spreader and $\boldsymbol{S_{\beta \tilde{\delta}}} = 0$ otherwise, then $P[\boldsymbol{S_{\beta \tilde{\delta}}} = 1] = X_{\beta \tilde{\delta}}$. In this way, the independence of random variables implies  that $P[\boldsymbol{S_{\beta \tilde{\delta}}} = 1, \boldsymbol{S_{\alpha \tilde{\gamma}}} = 1] = P[\boldsymbol{S_{\beta \tilde{\delta}}} = 1] P[\boldsymbol{S_{\alpha \tilde{\gamma}}} = 1] = X_{\beta \tilde{\delta}} X_{{\alpha \tilde{\gamma}}}$. Cator and Van Mieghem~\cite{Cator2014} proved rigorously that the states of any two nodes in the SIS model are
non-negatively correlated for all finite graphs. This result can be easily extended to our case, since we are considering constant rates and
Markovian processes. Due to the positive contribution of the infected nodes we have $P[\boldsymbol{S_{\beta \tilde{\delta}}} = 1 | \boldsymbol{S_{\alpha \tilde{\gamma}}} = 1] \geq P[\boldsymbol{S_{\beta \tilde{\delta}}} = 1]$, implying that the model is always overestimated. A similar conclusion was also obtained in~\cite{Mieghem09} for the monolayer case.

Naturally, the order parameter, also called macro-state variable, is defined as the average of the individual probabilities, formally given by
\begin{equation}
 \rho = \frac{1}{nm}X_{\beta \tilde{\delta}} U^{\beta \tilde{\delta}}.
\end{equation}
Note that the steady state is not an absorbing state in the Markov sense, since there is a set of possible states where the system remains trapped and there is a stochastic variation over time. In addition, note that there are many different configurations for which the fraction of infected nodes is the same. More formally, there is a set of states above the threshold, which have finite probability larger than zero, configuring a meta-state. The only absorbing state of this set of equations is thus the disease-free state, since when it is reached the (micro and macro) dynamics stops.

Furthermore, one of the most important concepts on disease spreading processes is the epidemic threshold: before the threshold, the system is in a disease-free state. On the other hand, when increasing the spreading rate it drives the population to an endemic state. In other words, there is a nonzero probability that the disease remains on the population, configuring the meta-state described above. Analogously to the results for monolayer systems we have a critical point given as 
\begin{equation} \label{eq:threshold}
 \left(\frac{\mu}{\lambda} \right)_c = \Lambda_1,
\end{equation}
where $\Lambda_1$ is the largest eigenvalue of $\mathcal{R}$. The complete derivation of the critical point is presented in Appendix~\ref{sec:app_sis_thr}. Observe that the eigen-structure of the tensor $\mathcal{R}$ is the same as for the matrix $R$ in~\cite{Cozzo2013}, since it can be understood as a flattened version of the tensor $\mathcal{R}_{\beta \tilde{\delta}}^{\alpha \tilde{\gamma}}(\lambda, \eta)$. As argued in~\cite{DeDomenico2013}, the supra-adjacency matrix
corresponds to a unique unfolding of the fourth-order tensor $\mathcal{R}$ yielding square matrices. Moreover, if $\eta M_{\nu \tilde{\delta}}^{\xi \tilde{\gamma}} E_{\xi}^{\nu}(\beta \beta) \ll \lambda M_{\beta \tilde{\gamma}}^{\alpha \tilde{\xi}} E_{\tilde{\xi}}^{\tilde{\gamma}}(\tilde{\delta}\tilde{\delta})$, the critical point is dominated by the individual layer behavior and the epidemic threshold is approximated to that of a SIS model on monolayers, when considering the union of $m$ disjoint networks. Consequently, the epidemic threshold is determined by the largest eigenvalue, considering all layers. The same conclusion was reached in~\cite{Cozzo2013} using perturbation theory on the supra-contact matrix. 

Finally, the nodal probability on the steady state can be bounded by
\begin{equation} \label{eq:bounds}
 1 - \frac{1}{1 + \frac{d_{\beta \tilde{\delta}}}{d^{\text{min}}} \left[ \left( \frac{\lambda}{\mu} \right) d^{\text{min}} - 1 \right]} \leq X^{\infty}_{\beta \tilde{\delta}} \leq 1 - \frac{1}{\left( \frac{\lambda}{\mu} \right) d_{\beta \tilde{\delta}} + 1},
\end{equation}
where $X^{\infty}_{\beta \tilde{\delta}}$ denotes the probability that node $\beta$ in layer $\tilde{\delta}$ is in the steady state regime, $d_{\beta \tilde{\delta}} = \mathcal{R}_{\beta \tilde{\delta}}^{\alpha \tilde{\gamma}}(\lambda, \eta)  U_{\alpha \tilde{\gamma}}$ (also defined in~\ref{eq:d_beta_delta}) and $d^{\text{min}} = \text{Min}\{d_{\beta \tilde{\delta}}\}$. The derivation of such bounds are shown in details on Appendix~\ref{sec:app_sis_bounds}.  Interestingly, observe that the higher $d^{\text{min}}$, the closer the lower and upper bounds. In the extreme case $\left( \frac{\lambda}{\mu} \right) \rightarrow \infty$ the bounds approach each other and all nodes tend to be infected. Phenomenologically, the latter parameter configuration models the limiting case of a SI-like scenario, where $\mu = 0$. In such a dynamical process all individuals are infected in the steady state.

\section{Spectral analysis of $\mathcal{R}(\lambda, \eta)$} \label{sec:Spectra}

As observed on the previous section, the supra adjacency tensor $\mathcal{R}(\lambda, \eta)$ plays a major role on the epidemic process. Consequently, a deeper analysis of the spectral properties of such object can give us further insights about the whole process. First of all, the generalization of the eigenvector problem to the eigentensor is described on Appendix~\ref{sec:app_eig_sec}, allowing us to use some well established linear algebra tools. Additionally, in this section we generalize the spectral results of interlacing, obtained in~\cite{Garcia2014,Cozzo2015}, to the tensorial description adopted here. Besides, we also make use of the inverse participation ratio, $\text{IPR}(\Lambda)$, as a measurement of eigenvalue localization~\cite{Goltsev2012}. As a convention, we assume that the eigenvalues are ordered as $\Lambda_1 \geq \Lambda_2 \geq ... \Lambda_{nm}$ and the individual layer eigenvalues are denoted as $\Lambda^l_{i}$.

\subsection{Interlacing properties} \label{sec:Spectra_interlacing}

\begin{figure*}[!t]
\begin{center}
\includegraphics[width=0.8\linewidth]{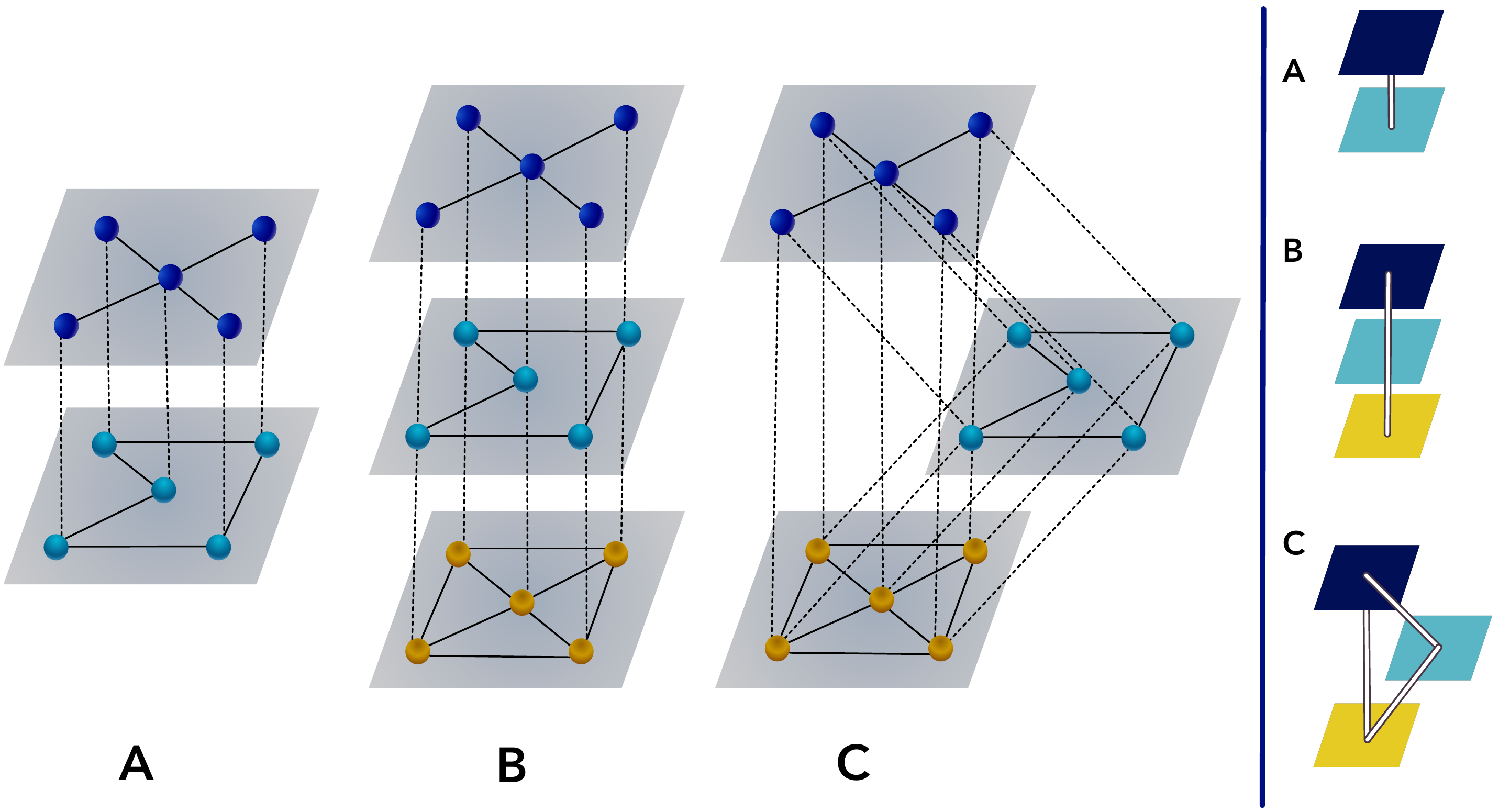}
\end{center}
\caption{Schematic Illustration of the 3 multilayer networks cases considered as examples. Top panels represent the original networks which give rise to three distinct configurations for the networks of layers. See the text for more details.}
\label{fig:schematic}
\end{figure*}

\begin{table}[t] 
\begin{center}
\caption{Structure and spectra of the normalized network of layers $\Phi_{\tilde{\delta}}^{\tilde{\gamma}}(\lambda, \eta)$. The eigenvalues assumes that the average degree of each layer, $\langle k^{l} \rangle$, is the same, i.e. $\langle k^{l}  \rangle = \langle k \rangle, \forall l$.}
\begin{tabular}{|c|c|c|}
\hline
Network & $\Phi_{\tilde{\delta}}^{\tilde{\gamma}}(\lambda, \eta)$ & Eigenvalues \\
\hline
\multirow{3}{*}{Line with 2 nodes} &

\multirow{3}{*}{$\begin{bmatrix}
 \langle k^{l = 1} \rangle & \frac{\eta}{\lambda} \\
 \frac{\eta}{\lambda} & \langle k^{l = 2} \rangle 
\end{bmatrix}$} & $\langle k \rangle - \frac{\eta}{\lambda}$\\
    & & $\langle k \rangle + \frac{\eta}{\lambda}$ \\
    & & \\
\hline
\multirow{4}{*}{Line with 3 nodes} &

\multirow{4}{*}{$\begin{bmatrix}
 \langle k^{l = 1} \rangle & \frac{\eta}{\lambda}  & 0 \\
 \frac{\eta}{\lambda} & \langle k^{l = 2} \rangle  & \frac{\eta}{\lambda} \\
 0 & \frac{\eta}{\lambda} & \langle k^{l = 3} \rangle
\end{bmatrix}$} & $\langle k \rangle$\\
 & & $\langle k \rangle - \sqrt{2} \frac{\eta}{\lambda}$ \\
 & & $\langle k \rangle + \sqrt{2} \frac{\eta}{\lambda}$ \\
 & & \\
\hline
\multirow{4}{*}{Multiplex} &

\multirow{4}{*}{$\begin{bmatrix}
 \langle k^{l = 1} \rangle & \frac{\eta}{\lambda}  & \frac{\eta}{\lambda} \\
 \frac{\eta}{\lambda} & \langle k^{l = 2} \rangle  & \frac{\eta}{\lambda} \\
 \frac{\eta}{\lambda} & \frac{\eta}{\lambda} & \langle k^{l = 3} \rangle
\end{bmatrix}$} &  $\langle k \rangle - \frac{\eta}{\lambda}$\\
& & $\langle k \rangle - \frac{\eta}{\lambda}$\\
& & $\langle k \rangle + 2 \frac{\eta}{\lambda}$\\
& & \\
\hline
\end{tabular}
\label{tab:spec_net_net}
\end{center} 
\end{table}

Invoking the unique mapping presented on Appendix~\ref{sec:app_eig_sec} and considering the results of~\cite{Garcia2014,Cozzo2015}, we can use the interlacing properties to relate the spectra of the multilayer network with the spectra of the network of layers. First of all, we define the normalized network of layers in terms of the supra contact tensor as 
\begin{equation} \label{eq:Net_net}
 \Phi_{\tilde{\delta}}^{\tilde{\gamma}}(\lambda, \eta) = \frac{1}{n} \mathcal{R}_{\beta \tilde{\delta}}^{\alpha \tilde{\gamma}}(\lambda, \eta) U^{\beta}_{\alpha},
\end{equation}
where we are implicitly assuming a multilayer network in which the layers have the same number of nodes and a dependency on the spreading rates (the demonstration that such tensor is an unfolding of the matrix exposed in~\cite{Garcia2014} is shown on Appendix~\ref{sec:Appendix_A}). Additionally, let's denote by $\mu_1 \geq \mu_2 \geq ... \geq \mu_{m}$ the ordered eigenvalues of $\Phi_{\tilde{\delta}}^{\tilde{\gamma}}(\lambda, \eta)$. Following ~\cite{Garcia2014}, the interlacing properties imply 
\begin{equation}
 \Lambda_{nm-m+j} \leq \mu_{j} \leq \Lambda_{j},
\end{equation}
for $j = m, ..., 1$. As examples, Table~\ref{tab:spec_net_net} shows the
spectrum of three simple networks of layers that can be computed
analytically: a line with two and three nodes and a triangle. Figure~\ref{fig:schematic} shows a schematic illustration of those 3 multilayer networks. 

Furthermore, using similar arguments we can also obtain results for the normalized projection, formally given as
\begin{equation} \label{eq:norm_agregated}
 \mathbf{P}_\beta^\alpha = \frac{1}{m} \mathcal{R}_{\beta \tilde{\delta}}^{\alpha \tilde{\gamma}}(\lambda, \eta) U^{\tilde{\delta}}_{\tilde{\gamma}},
\end{equation}
whose ordered eigenvalues, denoted by $\nu_1 \geq \nu_2 \geq ... \geq \nu_{m}$, also interlace with the supra contact tensor satisfying
\begin{equation}
 \Lambda_{nm-n+j} \leq \nu_{j} \leq \Lambda_{j},
\end{equation}
for $j = n, ..., 1$. Finally, the adjacency tensor of an extracted layer also interlaces, yielding
\begin{equation}
 \Lambda_{nm-n+j} \leq \Lambda_{j}^l \leq \Lambda_{j},
\end{equation}
for $j = n, ..., 1$. These results show that the eigenvalue of the
multilayer adjacency tensor is always larger than or equal to all of the
eigenvalues of the individual isolated layers as well as the network of
layers.

The interlacing properties presented here imply some constraints to the epidemic threshold. As advanced in~\cite{Garcia2014}, let $\Lambda_i(\mathcal{M})$ be the $i$-th eigenvalue of the tensor $\mathcal{M}$ and consider that the set of eigenvalues is ordered as before. Moreover, for simplicity, we suppress the argument when referring to the supra-contact matrix. First of all, assuming a fixed ratio of spreading rates, we observe that the eigenvalue of the multilayer follows
\begin{equation}
 \left(\frac{\lambda}{\mu} \right)_c^{\tilde{r}} = \frac{1}{\Lambda_1(A^{\alpha}_{\beta}(\tilde{r}))} \geq \frac{1}{\Lambda_1}, \hspace{1cm} \forall \tilde{r} \in 1, 2, ..., m,
\end{equation}
where $\left(\frac{\lambda}{\mu} \right)_c^{\tilde{r}}$ is the critical point for the single layer $\tilde{r}$ and
\begin{equation}
 \left(\frac{\lambda}{\mu} \right)_c^{\Phi} = \frac{1}{\Lambda_1(\Phi_{\tilde{\delta}}^{\tilde{\gamma}})} \geq \frac{1}{\Lambda_1},
\end{equation}
where $\left(\frac{\lambda}{\mu} \right)_c^{\Phi}$ denotes the critical point of the network of layers. Finally, considering the projection, we get
\begin{equation}
 \left(\frac{\lambda}{\mu} \right)_c^{\mathbf{P}} = \frac{1}{\Lambda_1(\mathbf{P}_\beta^\alpha)} \geq \frac{1}{\Lambda_1},
\end{equation}
where $\left(\frac{\lambda}{\mu} \right)_c^{\mathbf{P}}$ is the critical
point of the normalized projection. Thus, the spreading process on the
whole system is at least as efficient as it is on the layers and on the
network of layers. Note that efficiency is understood here in terms of
the position of the critical point, and not regarding the fraction of
infected individuals in the steady state.

\subsection{Localization and spreading of diseases}

Next, we investigate the behavior of the system near the phase transition and whether the phenomenon of disease localization shows up. These two issues were explored for monoplex networks in~\cite{Mieghem2012} and~\cite{Goltsev2012}, respectively, but have not been addressed for the case of multilayer systems. The nodal probabilities can be written as a linear combination of the eigenbasis of $\mathcal{R}$ as
\begin{equation}
 X_{\beta \tilde{\delta}} = \sum_{\Lambda} c(\Lambda) f_{\beta \tilde{\delta}}(\Lambda),
\end{equation}
where $c(\Lambda)$ are the projections of $X_{\beta \tilde{\delta}}$ on the eigentensors $f$. Similarly to~\cite{Goltsev2012}, substituting such expression on the middle term of eq.~\ref{eq:steady} we obtain
\begin{equation}
c(\Lambda) = \sum_{\alpha \tilde{\gamma}} \frac{\lambda \sum_{\Lambda'}  c(\Lambda') \Lambda' f_{\alpha \tilde{\gamma}}(\Lambda') f_{\alpha \tilde{\gamma}}(\Lambda) }{\lambda \sum_{\Lambda'} c(\Lambda') \Lambda' f_{\alpha \tilde{\gamma}}(\Lambda') + \mu}.
\end{equation}

Considering only the contributions of the first eigenvalue and eigentensor, for $\lambda \geq \lambda_c$, the first order approximation of the macro state parameter is $\rho \approx \alpha_1 \tau$, where $\tau = \left( \frac{\lambda}{\mu} \Lambda_1 - 1 \right)$, which yields
\begin{equation}
 \alpha_1 = \frac{f_{\beta \tilde{\delta}}(\Lambda_1) U^{\beta \tilde{\delta}}}{n m (f_{\beta \tilde{\delta}}(\Lambda_1))^3 U^{\beta \tilde{\delta}}}.
\end{equation}
Such an expression is exact if there is a gap between the first two eigenvalues~\cite{Mieghem2012, Goltsev2012}. Furthermore, considering two eigentensors we have $\rho \approx \alpha_1 \tau + \alpha_2 \tau^2$. Besides, following a similar approach as in~\cite{Goltsev2012} we can use the inverse participation ratio: 
\begin{equation} \label{eq:ipr}
 \text{IPR}(\Lambda) \equiv \left( f_{\beta \tilde{\delta}}(\Lambda) \right)^4 U^{\beta \tilde{\delta}}.
\end{equation}
In the limit of $nm \rightarrow \infty$, if the $\text{IPR}(\Lambda)$ is of order $\mathcal{O}(1)$ the eigentensor is localized and the components of $f_{\beta \tilde{\delta}}(\Lambda)$ are of order $\mathcal{O}(1)$ only for a few nodes. On the other hand, if  $\text{IPR}(\Lambda) \rightarrow 0$ then this state is delocalized and the components of $f_{\beta \tilde{\delta}}(\Lambda) \sim \mathcal{O}\left(\frac{1}{\sqrt{nm}}\right)$. Additionally, another possible scenario, completely different from the traditional single layer one, is possible if we consider localization on layers instead of on a fraction of nodes. In such a case, the $IPR(\Lambda)$ will be of order $\mathcal{O}(1/n)$ in the localized phase, whereas it will be of order $\mathcal{O}(1/nm)$ in the delocalized phase. This is because, in the localized phase the components of the eigentensor are of order $\mathcal{O}(1/\sqrt{n})$  for all the nodes in the dominant layer and of order zero for nodes in other layers. Observing that, one easily realizes that the correct finite-size scaling to take in order to characterize such a transition is $m\rightarrow\infty$, i.e., the number of layers goes to infinity while the number of nodes per layer remains constant. In fact, in this limit $IPR(\Lambda)$ will vanish on one side of the transition point while remaining finite on the other side. In this way, we can observe localized states also in the case in which there is no possibility for localization in each of the layers if they were isolated.

\section{Monte Carlo simulations}

We next compare the analytical results with Monte Carlo simulations of the spreading process. The method proposed in~\cite{Ferreira2012, Mata2015} is adapted here to the case of multilayer networks. At each time step the time is incremented by $\Delta t = \frac{1}{(\mu N_i + \lambda N_k + \eta N_m)}$, where $N_i$ is the number of infected nodes, and $N_k$ and $N_m$ are the number of intra-layer and inter-layer edges emanating from them, respectively. With probability $\frac{\mu N_i}{(\mu N_i + \lambda N_k + \eta N_m)}$, one randomly chosen infected individual becomes susceptible. On the other hand, with probability $\frac{\lambda N_k}{(\mu N_i + \lambda N_k + \eta N_m)}$, one infected individual, chosen with a probability proportional to its intra-layer degree, spreads the disease to an edge chosen uniformly random. Finally, with probability $\frac{\eta N_m}{(\mu N_i + \lambda N_k + \eta N_m)}$ one infected individual, chosen with a probability proportional to its inter-layer degree, propagates the disease to an edge chosen uniformly. If an edge between two infected individuals is selected during the spreading, nothing happens, only time is incremented. The process is iterated following this set of rules, simulating the continuous process described by the SIS scenario.

The quasi-stationary state (QS)  method~\cite{Ferreira2012, Mata2015} restricts the dynamics to non-absorbing states. Every time the process tries to visit an absorbing state, it is substituted by an active configuration previously visited and is stored on a list with $M$ configurations, constantly updated. With a probability $p_r$ a random configuration on such a list is replaced by the actual configuration. In order to extract meaningful statistics from the quasi-static distribution, denoted by $\bar{P}(n^I)$, where $n^I$ is the number of infected individuals, the system must be on the stationary state and a large number of samples must be extracted. In this way we let the simulations run during a relaxation time $t_r$ and extract the distribution $\bar{P}(n^I)$ during a sampling time $t_a$. The threshold can be estimated using the modified susceptibility~\cite{Ferreira2012}, given by
\begin{equation}
 \chi = \frac{ \langle (n^I)^2 \rangle - \langle n^I \rangle^2 }{\langle n^I \rangle} = nm \left( \frac{ \langle (\rho^{QS})^2 \rangle - \langle \rho^{QS} \rangle^2 }{\langle \rho^{QS} \rangle} \right),
\end{equation}
where $\rho^{QS}$ is the quasi-stationary distribution
$\bar{P}(n^I)$. As argued in~\cite{Ferreira2012, Mata2015} the
susceptibility presents a peak at the phase transition on finite
systems. Such measure is the coefficient of variation of the temporal
distribution of states over time on the steady state. Note that the
magnitude of the susceptibility $\chi$ is not of primary interest to us,
but rather the position of its maximum value with respect to
$\mu/\lambda$, since it will coincide with the critical threshold for
sufficiently large systems.

In addition, after obtaining the curves of $\chi \times \lambda$ by the
QS method, we also apply a moving average filter in order to get rid of
the noise. Such an approach improves the visual quality of the plots and
does not interfere on the results, since the order of
magnitude of the noise is smaller than those of the peaks corresponding to the transition points.

The parameters used in the QS method are $p_r = 0.01$, $t_a$ varies from $10^{5}$ to $10^{6}$ and $t_r$ varies from $10^{5}$ to $3 \times 10^{6}$ in order to obtain a smoother curve. The QS method demands a large sample size, since it is estimating the variance of a distribution. Moreover, we construct the $\chi \times \lambda$ curves in steps of $\Delta \lambda = 10^{-3}$ and the moving average window has 5 points.

\section{2-Layer multiplex systems}

In this section we numerically study 2-layer multiplex systems. First, we focus on the phase diagram of the spreading process as a function of the inter and intra layer spreading rates for both, SIS and SIR scenarios. Next, we analyze the spectral properties of such systems, comparing with results of Section~\ref{sec:Spectra}. Finally, we perform Monte Carlo simulations that show the existence of multiple susceptibility peaks on multiplex networks. The latter results are analyzed in terms of the spectral properties of $\mathcal{R}(\lambda, \eta)$.

\subsection{Numerical solution} \label{sec:2L_numerical}

\begin{figure}[!t]
\begin{center}
\subfigure[][SIS]{\includegraphics[width=0.98\linewidth]{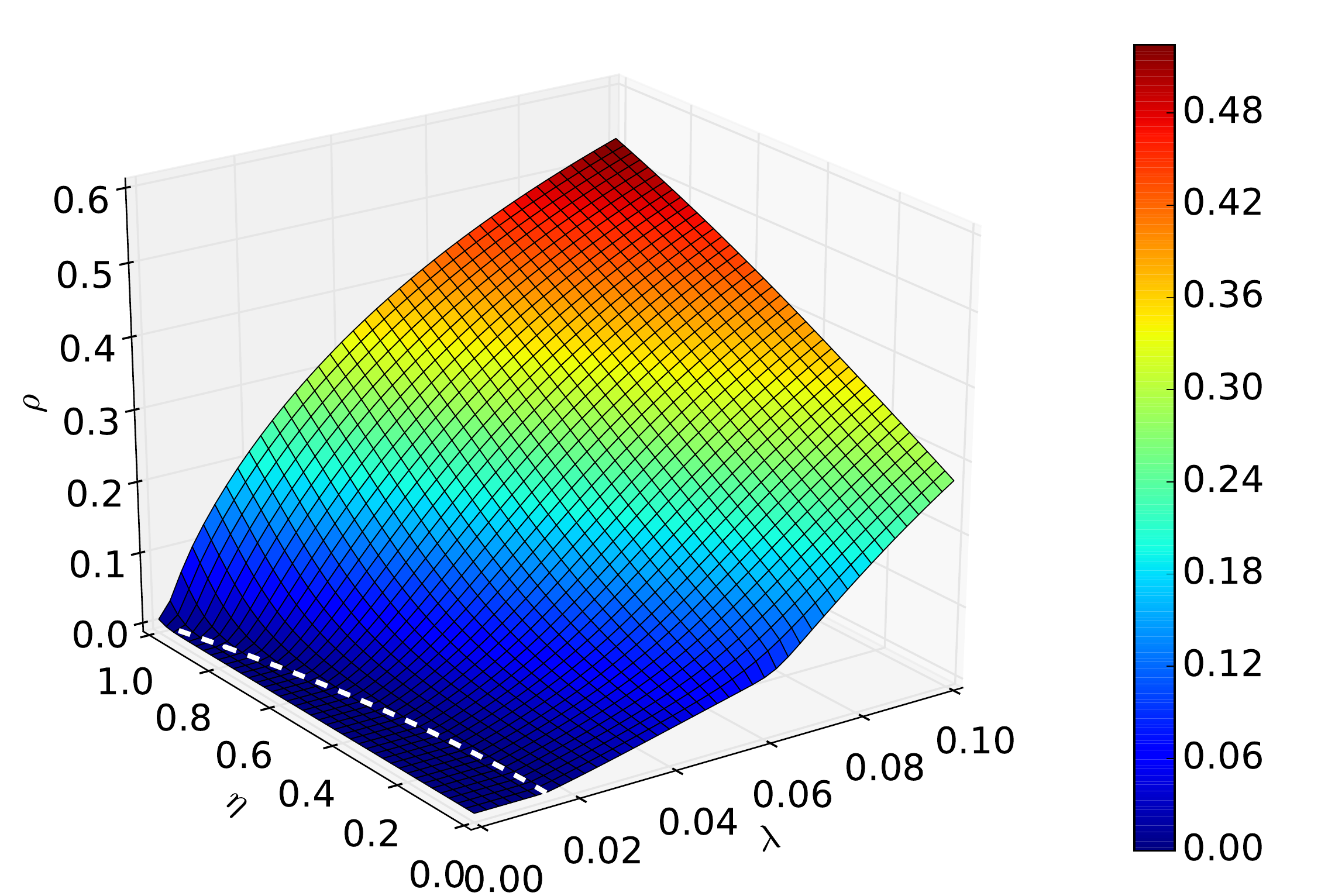}}
\subfigure[][SIR]{\includegraphics[width=0.98\linewidth]{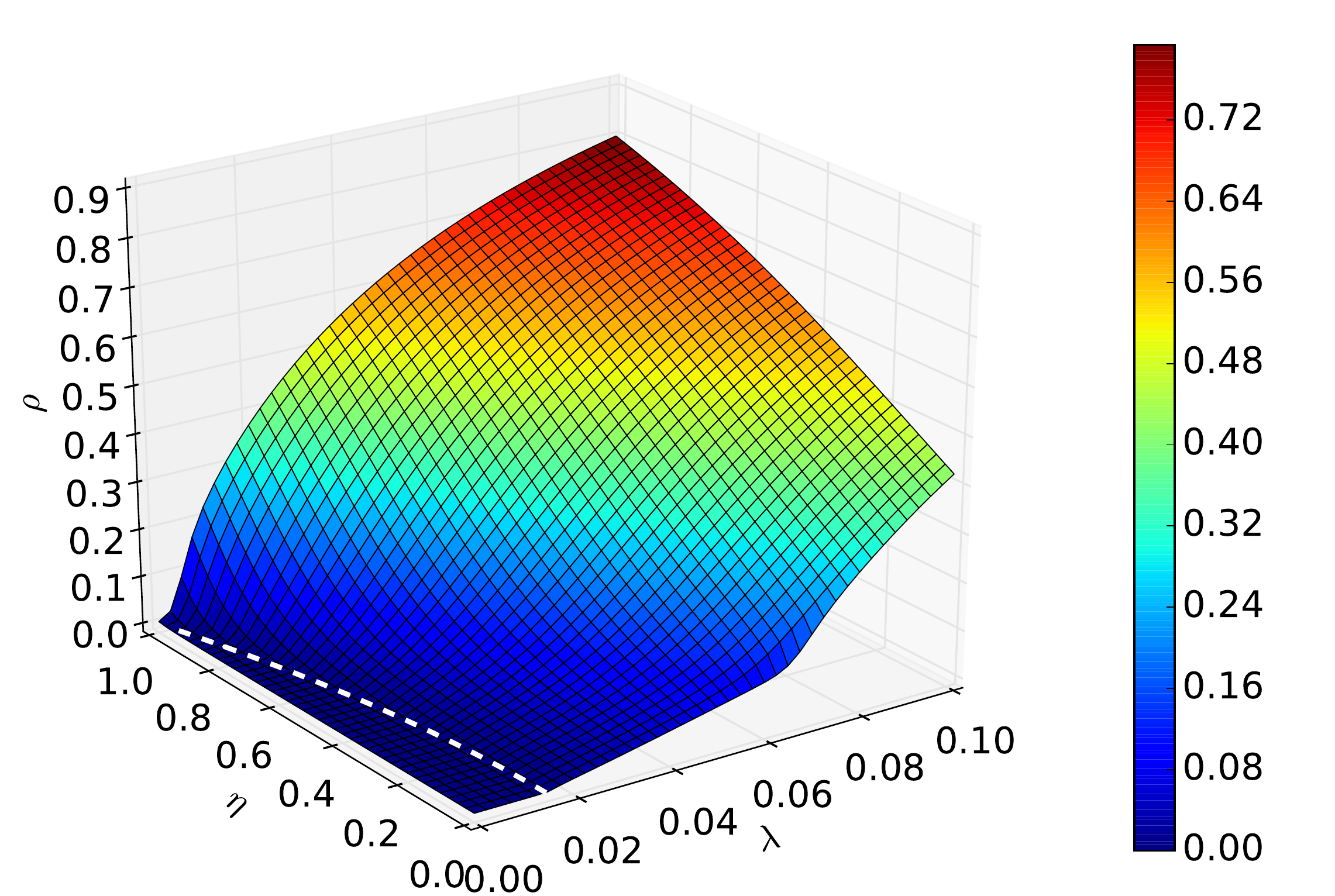}}
\end{center}
\caption{Phase diagrams over a 2-Layer multiplex system, where each layer is a scale-free network with  $n=10^4$ nodes, for a fixed value of $\mu=1$. (a) Density of spreaders as a function of the parameters $\eta$ and $\lambda$. (b) Density of recovered individuals as a function of the parameters $\eta$ and $\lambda$. Colors represent the fraction of spreaders and the white line is the threshold calculated using equation~\ref{eq:threshold}.}
\label{fig:Phase}
\end{figure}

\begin{figure*}[!t]
\begin{center}
\subfigure[][Global]{\includegraphics[width=0.32\linewidth]{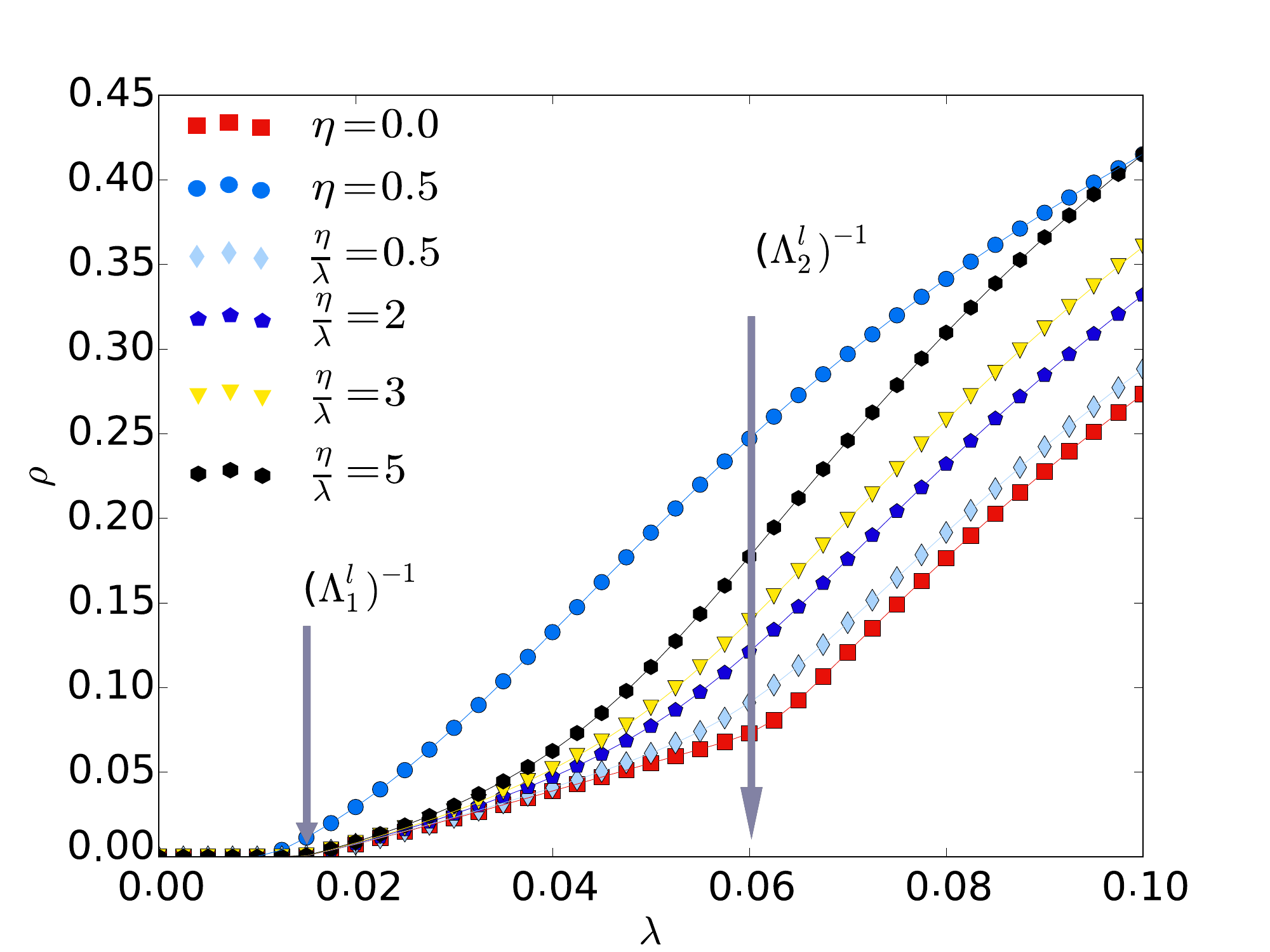}}
\subfigure[][First layer]{\includegraphics[width=0.32\linewidth]{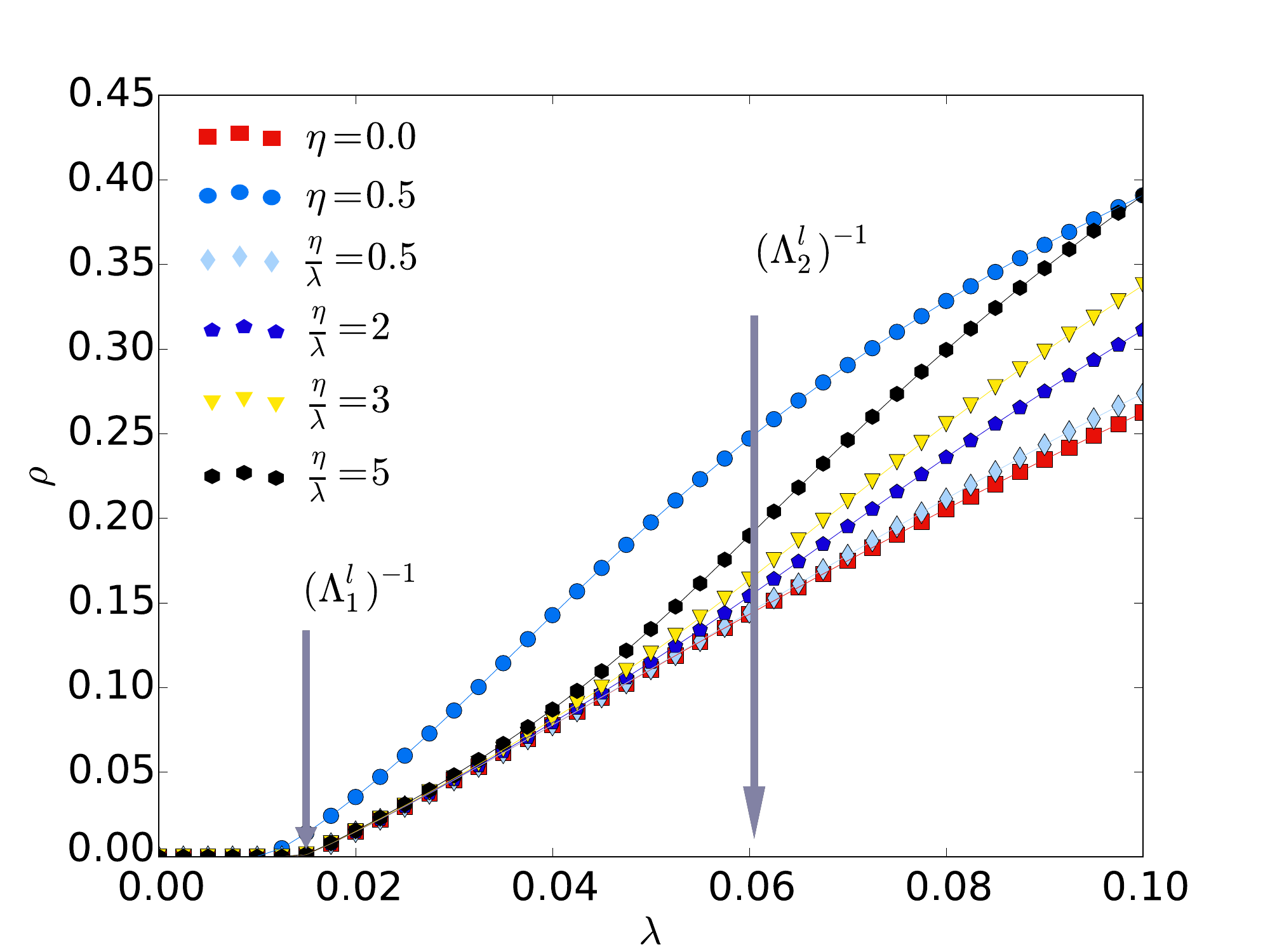}}
\subfigure[][Second layer]{\includegraphics[width=0.32\linewidth]{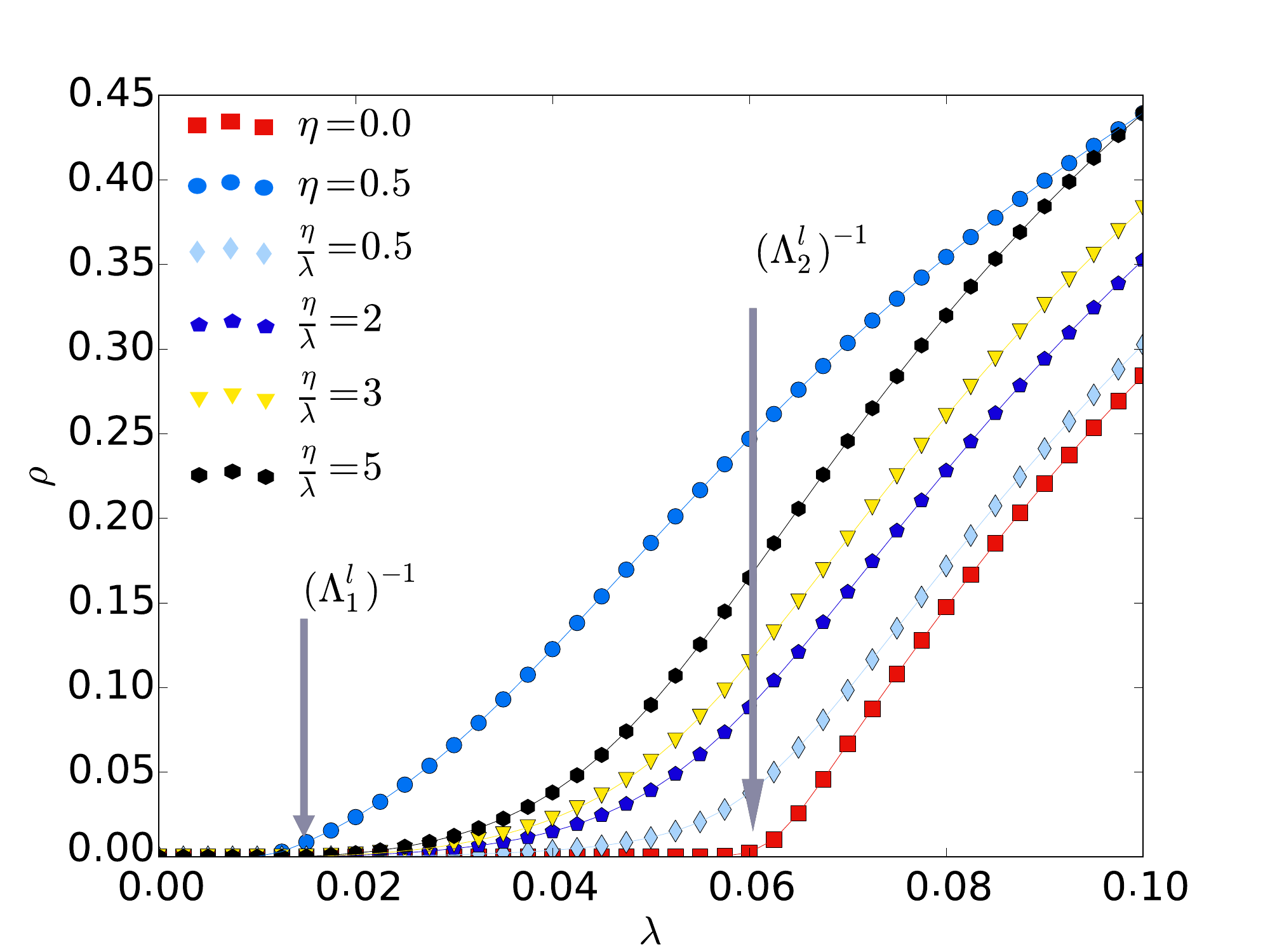}}
\end{center}
\caption{Individual layer behavior over a 2-Layer multiplex system. Each layer has  $n=10^4$ for a fixed value of $\mu=1$. The results considering both layers are shown in (a), while the dynamics in the individual layers are shown in (b) ($P(k) \sim k^{-2.5}$) and (c) ($P(k) \sim k^{-4.5}$). The arrows indicate the layers leading eigenvalues.}
\label{fig:PhaseLayer}
\end{figure*}

Results shown in this section are the numerical solutions of the ODE systems~\ref{eq:SIS_tensor} (SIS) and~\ref{eq:SIR_tensor} (SIR) using a Runge-Kutta (4,5) algorithm~\cite{Dormand1980}. We consider a 2 layer multiplex network ($m = 2$), where each layer has $n = 10^4$ nodes. In order to build a multiplex network where the epidemic thresholds associated to the individual layers are well separated, we must guarantee that $\Lambda_1^l \gg \Lambda_2^l$. Therefore, we chose the degree distribution of the first layer to be $P(k) \sim k^{-2.5}$, whereas that of the second layer is $P(k) \sim k^{-4.5}$. Both layers are created using the uncorrelated configuration model~\cite{Viger2005}. Moreover, we consider a multilayer network in which every node has its counterpart on the other layer. This pairing of nodes of different layers is made randomly. Each result is the solution considering one single (and fixed) multiplex network.

Figure~\ref{fig:Phase} shows the phase diagram considering the average fraction of spreaders for the SIS dynamics (or recovered for the SIR dynamics) as the macro-state variable as a function of the spreading parameters $\lambda$ and $\eta$ for a given recovering rate $\mu = 1$. The dashed white line denotes the epidemic threshold obtained from eq.~\ref{eq:threshold}. In (a) we show the SIS scenario, while (b) corresponds to the SIR model. In both cases, it is possible to observe two changes on the system's behavior. The first on the epidemic threshold, while the second near the epidemic threshold of the second layer. In addition, we note the agreement between the theoretical epidemic thresholds and the numerical results. Furthermore, the higher $\eta$, the lower the epidemic threshold, which is a consequence of the eigentensor problem. Also note that $\rho$ increases for a fixed $\lambda$ as $\eta$ increases, even for $\lambda \sim 0$, which means that in such extreme cases, the disease spreads mainly on the interlayer edges. 

Figure~\ref{fig:PhaseLayer} shows the phase diagram for $\mu = 1$ and different values of the parameter $\eta$ for the SIS dynamics. For $\eta = 0$ we have no inter-layer spreading, while for $\eta = 0.5$ we have a fixed spreading rate, independent of the intra-layer rates. In addition, we also evaluated cases where the ratio $\frac{\eta}{\lambda}$ is constant. In Fig.~\ref{fig:PhaseLayer} (a) we have the global behavior of the system, which is an average of the individual behavior of the layers, represented in panels (b) and (c), since both layers have the same number of nodes. Furthermore, we also observe that the two individual networks show different behaviors near the epidemic threshold~\cite{Mieghem2012}. The first layer (Fig.~\ref{fig:PhaseLayer} (b)) has a lower epidemic threshold than the second. However $\rho$ grows (as a function of $\lambda$) slower than in the second. This feature can be observed clearly in Fig.~\ref{fig:PhaseLayer} (b) and (c), where we show results for $\eta = 0$, that is, when there is no spreading between the layers. 

Considering the discrete system, Cozzo et al.~\cite{Cozzo2013} verified the shifting on the dominated layer (the largest amongst all individual eigenvalues) as the ratio $\frac{\eta}{\lambda}$ increases. Here we observe the same effect, as can be seen in Fig.~\ref{fig:PhaseLayer} (c). Additionally, we can also note another global change approximately beyond $\lambda > (\Lambda_2^l)^{-1}$. Our findings suggest the possibility of multiple phase transitions due to the multiplex structure of the network. It is noteworthy that in spite of the similarities between our continuous model and the discrete model~\cite{Cozzo2013}, both represent slightly different processes. On the continuous case, two events cannot happen at the same time. On the other hand, on the discrete model, every node contacts its neighbors on one discrete time step. Despite these differences, the results show that both the continuous and discrete formulations are phenomenologically similar.

\subsection{Spectral analysis} \label{sec:different_L}

\begin{figure}[!t]
\begin{center}
\includegraphics[width=1\linewidth]{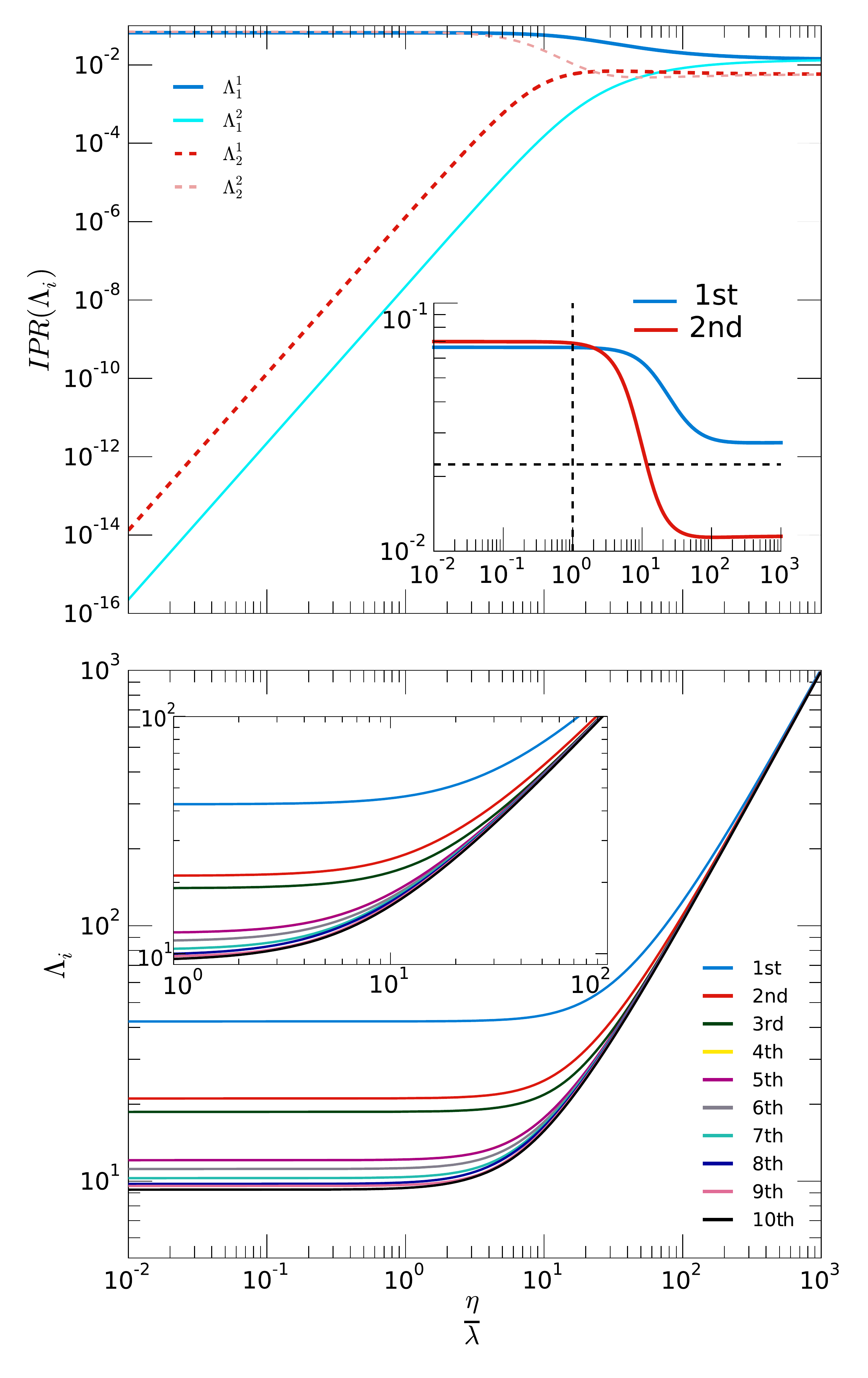}
\end{center}
\caption{Spectral properties of the tensor $\mathcal{R}(\lambda, \eta)$ as a function of the ratio $\frac{\eta}{\lambda}$ for a multiplex with two layers, the first with $\gamma \approx 2.2$, while the second $\gamma \approx 2.8$. Both have $\langle k \rangle \approx 8$. On the top panel we present the inverse participation ratio ($\text{IPR}(\Lambda)$) of the two larger eigenvalues and the individual layer contributions, while on bottom panel we show the leading eigenvalues. Every curve is composed by $10^3$ log spaced points, in order to have enough resolution.}
\label{fig:Localization_2}
\end{figure}

\begin{figure}[!t]
\begin{center}
\includegraphics[width=1\linewidth]{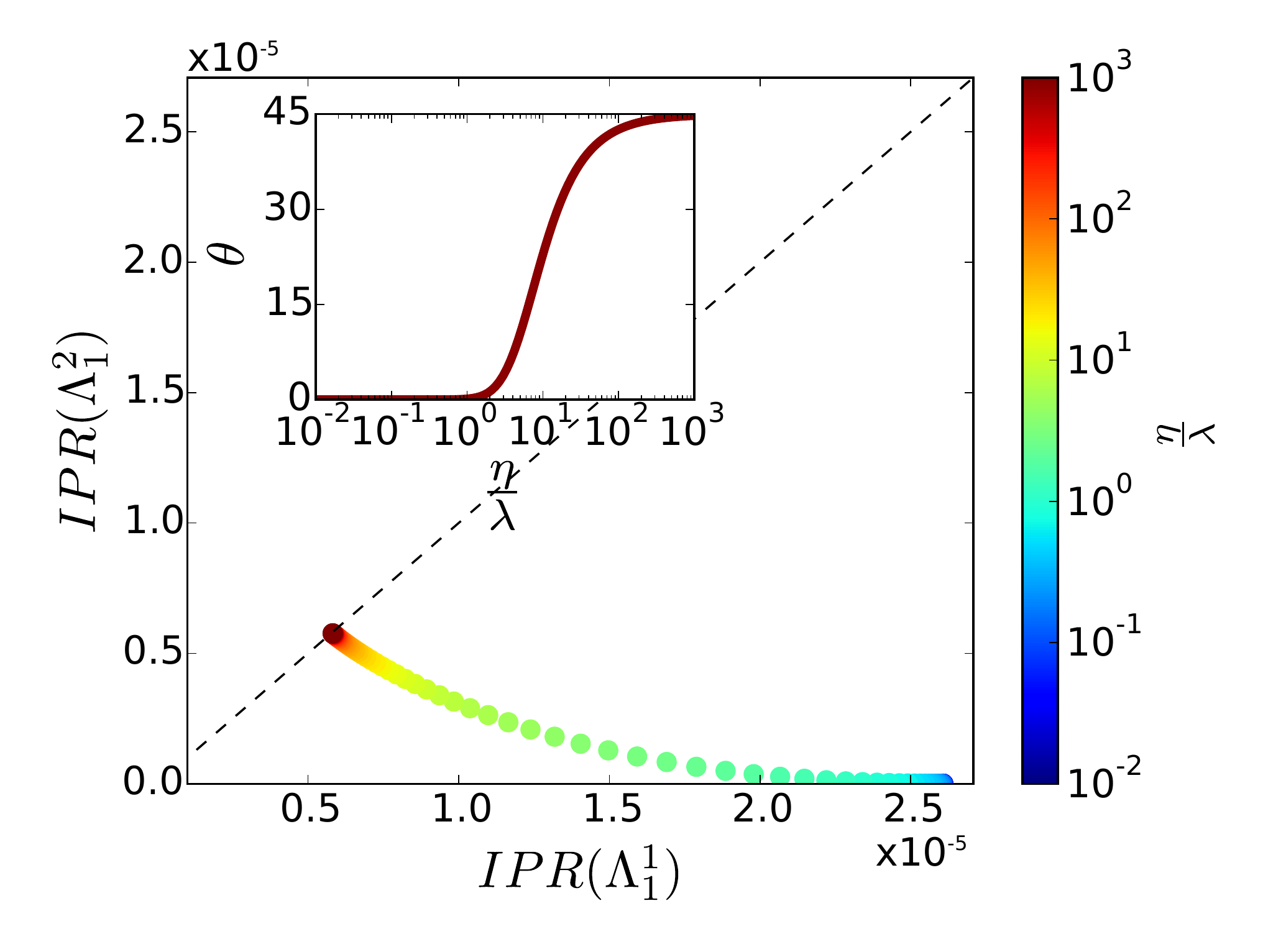}
\end{center}
\caption{Diagram of the contribution of each layer to the $\text{IPR}(\Lambda)$ for different values of the spreading ratio $\frac{\eta}{\lambda}$. The dashed line represents the case where both layers have the same contribution, i.e. a line with slope one. In the inset, we show the angle $\theta$ between the vector composed by the contributions of each layer to the $\text{IPR}(\Lambda)$, $v = \left[ \text{IPR}(\Lambda^1_1), \text{IPR}(\Lambda_1^2)\right]^T$, and the $x$-axis. The multiplex network used here is composed of two Erd\"os-R\'enyi networks, both with $n=5\times10^4$, the first layer $\langle k \rangle = 16$ ($(\Lambda_1^1)^{-1} \approx 0.0625 $), while the second $\langle k \rangle = 12$ ($(\Lambda_1^2)^{-1} \approx 0.0833$).}
\label{fig:IPR_vs_IPR}
\end{figure}

Since the epidemic process is described through the supra adjacency tensor $\mathcal{R}(\lambda, \eta)$, its spectral properties give us some insights about the whole process, especially about the critical properties of the systems under analysis. In this section we focus on the spectral analysis of such tensor as a function of the ratio $\frac{\eta}{\lambda}$ considering a 2-layer multiplex network with two different layers, i.e., there is a distance between the leading eigenvalues of each layer. Some important aspects of the spectral properties are left to Appendix~\ref{sec:app_2L}, where we present an analytical approach to the problem of eigenvalue crossings on Appendix~\ref{sec:app_crossing}. We focus on two special cases in increasing order of complexity: (i) the identical case, presented on Appendix~\ref{sec:app_identical}, where both layers are exactly the same $-$ i.e., there is a high correlation between the degree on each layer $-$; and (ii) the non-identical case, discussed in Appendix~\ref{sec:app_similar}, where both layers have the same degree distribution, but different configurations. 

In this section we focus on the case of two different layer structures, with spaced leading eigenvalues. Considering a multiplex network made up of two scale-free networks with $\gamma \approx 2.2$ and $\gamma \approx 2.8$. Both layers have $\langle k \rangle \approx 8$ and $n = 10^3$ nodes on each layer and the leading eigenvalues are $\Lambda^1_1 = 42.64$ for the first and $\Lambda^2_1 = 21.29$ for the second.

Figure~\ref{fig:Localization_2} shows the spectral properties of the tensor $\mathcal{R}(\lambda, \eta)$ as a function of the ratio $\frac{\eta}{\lambda}$. In contrast to the identical layers (see Appendix~\ref{sec:app_identical}) and the case of statistically equivalent layers (Appendix~\ref{sec:app_similar}), figures~\ref{fig:Localization_Identical} and~\ref{fig:Localization_Similar}, where some eigenvalues increase while others decrease, here all the observed eigenvalues always increase. Moreover, we do not observe any crossing or near-crossing behavior. Regarding $\text{IPR}(\Lambda)$, the same pattern as for the similar case is found: for small values of $\frac{\eta}{\lambda}$ and considering the first eigenvalue, the system appears localized on the first layer and delocalized on the second, while for $\text{IPR}(\Lambda_2)$, it is the contrary. For larger values of $\frac{\eta}{\lambda}$, both layers contribute equally to the $\text{IPR}(\Lambda)$. Furthermore, the main difference we observe for the current setup with respect to the two similar networks (see Fig.~\ref{fig:Localization_Similar}, presented on Appendix~\ref{sec:app_2L}), is that now no drastic change on the inverse participation ratio is found, as expected, since there is no near-crossing.

From figure~\ref{fig:Localization_2} we can also extract an important numerical result regarding the perturbation theory. We observed that in our case, considering a two spaced-individual layer eigenvalues problem, the leading eigenvalue can be approximated by the largest leading eigenvalue of the individual layers for $\frac{\eta}{\lambda} \lesssim 1$, such approximation becomes poorer as $\frac{\eta}{\lambda}$ increases, but it can be acceptable up to $\frac{\eta}{\lambda} \lesssim 10$, within a certain error. Apart from that, note that both eigenvalues tend to increase, while its difference tends to decrease.

Furthermore, analyzing the eigenfunction properties, Fig.~\ref{fig:IPR_vs_IPR} shows the contribution of each layer to the $\text{IPR}(\Lambda)$ considering different values of $\frac{\eta}{\lambda}$. Results correspond to a multiplex network composed by two Erd\"os-R\'enyi networks, both with $n=5\times10^4$, the first layer with $\langle k \rangle = 16$, while the second has $\langle k \rangle = 12$. Observe that for lower values of $\frac{\eta}{\lambda}$ the main contribution comes from one layer, configuring a localized state and consequently placed on one axis (the $x$-axis) of Fig.~\ref{fig:IPR_vs_IPR}. Then, when the ratio $\frac{\eta}{\lambda}$ increases, there is a transition to a delocalized state. This corresponds to an increase of the inverse participation ratio of the second layer, however at the expense of decreasing the value of the inverse participation ratio of the first layer. In other words, in the localized phase, only the entries of the eigenvector associated to the dominant layer are effectively populated, while the entries associated to other layers are not. In the delocalized phase all the entries are equally populated. The inset of the figure further evidences this transition: it represents the angle, $\theta$, between the vector composed by the IPR contributions, $v = \left[ \text{IPR}(\Lambda^1_1), \text{IPR}(\Lambda_1^2)\right]^T$, and the x-axis, where a change from zero to 45 degrees is observed as the ratio $\frac{\eta}{\lambda}$ is increased and the system goes from a localized to a delocalized state.

\subsection{Multiple susceptibility peaks} \label{sec:multiple_2Mux}

Mata and Ferreira showed that it is possible to have multiple susceptibility peaks on monoplex networks~\cite{Mata2015}. They studied the behavior of a SIS model on networks with $\gamma > 3$. Here we show that such phenomena also appear, in a natural way, on multilayer networks. Motivated by the findings reported in the latter sections, especially by the presence of a second change in the slope of $\rho$ as observed in figures~\ref{fig:Phase} and~\ref{fig:PhaseLayer}, we have performed extensive Monte Carlo simulations using the QS-method with the aim of determining as accurately as possible the points at which the transitions takes place for a 2-layer multiplex network. Here we use the multiplex built up in Section~\ref{sec:different_L}, since the leading eigenvalues of each layer are spaced. Note that our numerical simulations are performed on a fixed network, since we follow the quenched formalism. 

\begin{figure}[!t]
\begin{center}
\includegraphics[width=0.98\linewidth]{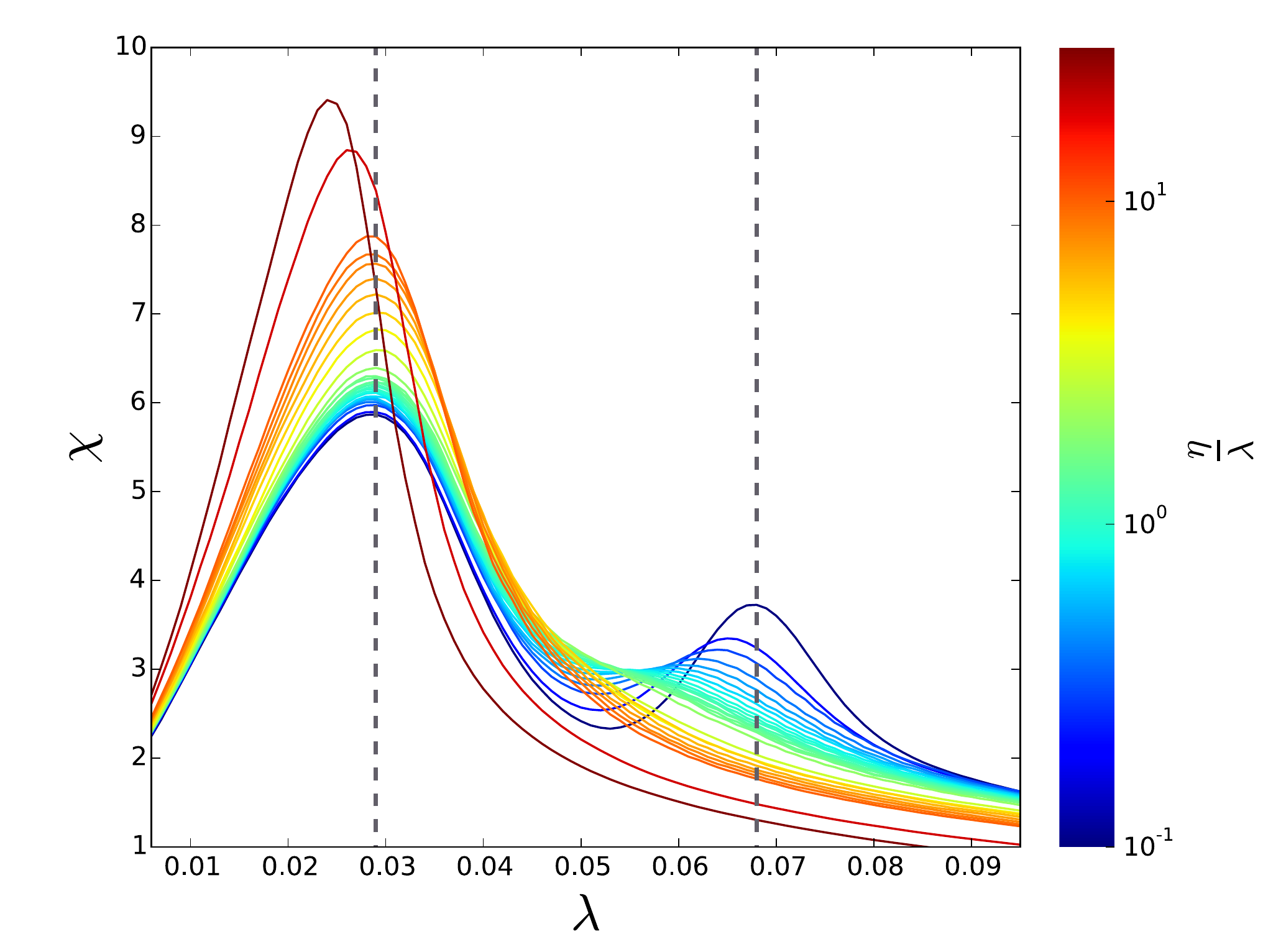}
\end{center}
\caption{Susceptibility, $\chi$, as a function of the spreading rate $\lambda$ for different ratios of inter and intra-layer spreading ratings, $\frac{\eta}{\lambda}$ for a fixed value of $\mu=1$ over a 2-Layer multiplex system, where each layer have  $n=10^3$, the first with $\gamma \approx 2.2$, while the second $\gamma \approx 2.8$. Both have $\langle k \rangle \approx 8$. The simulated values are $\frac{\eta}{\lambda} = $ 0.1, 0.2, 0.3, 0.4, 0.5, 0.6, 0.7, 0.8, 0.9, 1.0, 1.1, 1.2, 1.3, 1.4, 1.5, 1.6, 2, 3, 4, 5, 6, 7, 8, 9, 10, 20, 30.}
\label{fig:QS}
\end{figure}

Figure~\ref{fig:QS} shows that for low values of the ratio $\frac{\eta}{\lambda}$, both networks are weakly coupled and the system exhibits two well-defined susceptibility peaks (vertical dotted lines). However, as this ratio increases the peak signaling the presence of the second critical point decreases and eventually vanishes. In our simulations, we have observed that up to $\frac{\eta}{\lambda} \approx 1$, the second peak, although less defined, is still present. Beyond the latter point, only one peak remains. As $\frac{\eta}{\lambda}$ further increases, the position of the critical point remains the same, and the peak is even more well defined. Interestingly enough, if the ratio $\frac{\eta}{\lambda}$ continues to increase --- in our case beyond $\frac{\eta}{\lambda} \gtrsim 10$ --- the critical point shift to the left to values that are even smaller than the smallest critical point of the individual layers. It is worth highlighting that a similar qualitative behavior can be seen in the results shown in Fig.~\ref{fig:Phase} (a), where one can also observe a second change in the slope of $\rho$ near the leading eigenvalue of the second layer. This change also vanishes as the intra-layer spreading increases. 

Since the tensor $\mathcal{R}(\lambda, \eta)$ plays a major role on the spreading process, our spectral results can help understanding the observed critical dynamics. In epidemiological terms --- or in general for contagion processes ---, the localization of the disease on a certain layer means that most of the spreading is expected to take place on the nodes of that layer. Moreover, in addition to the localization on the layers, one can also have localization effects on specific nodes or groups of nodes, for instance.

In order to analytically explain this phenomenon, we evaluate $\text{IPR}(\Lambda)$ for the two leading eigenvalues, as this measure indicates the localization of an eigenstate, see Section~\ref{sec:different_L} (results shown in Fig.~\ref{fig:Localization_2}). Comparing the susceptibility and $\text{IPR}(\Lambda)$, we observe that $\text{IPR}(\Lambda_2)$ starts decaying for $\frac{\eta}{\lambda} \approx 1$ and crosses the value $\frac{1}{\sqrt{nm}}$, at which the associated eigenvector delocalizes, for $\frac{\eta}{\lambda} \approx 10$, comparing well with the point at which the second peak in the susceptibility decays and finally disappears. Moreover, $\text{IPR}(\Lambda_1)$ decays from $3 \lesssim \frac{\eta}{\lambda} \lesssim 10$, which coincides with the range where the remaining maximum in the susceptibility reaches higher values and is better defined. More interestingly, note that $\text{IPR}(\Lambda_1)$ is mainly composed by the contributions of the first layer for a lower spreading ratio, suggesting that it is localized on such layer. Therefore, our results suggest that the $\text{IPR}(\Lambda)$ is a proper measure to detect and predict the observed localization phenomena 
and potentially for $m$ localization transitions, as we will show on Section~\ref{sec:3L}. 

Regarding the definition of a critical point it is important to highlight that the concept of phase transition only applies in the infinite size limit (the thermodynamic limit). However, on the literature of complex network dynamics, specially for epidemic spreading, it is usual to use the terms critical point and phase transition on finite systems, since we find a behavioral change on that point. More importantly, for scale-free networks such point vanishes in the thermodynamic limit. Following the usual convention on the complex network literature, the first susceptibility peak observed on all the experiments can be classified as a critical point of a phase transition. On such point, the dynamics goes from a disease-free state to an endemic state. On the other hand, the second susceptibility peak cannot receive this classification, since the process is already on a endemic state. Although it cannot be considered as a critical point, we have a transition from a localized state to a delocalized state. In other words, before the second susceptibility peak most of the events take place on only one layer (the one with largest individual eigenvalue), while after this point both layers are active and spreading the disease.

\subsection{Second susceptibility peak analysis: Erd\"os-R\'enyi layers} \label{sec:sec_peak}

\begin{figure*}[!t]
\begin{center}
\includegraphics[width=0.98\linewidth]{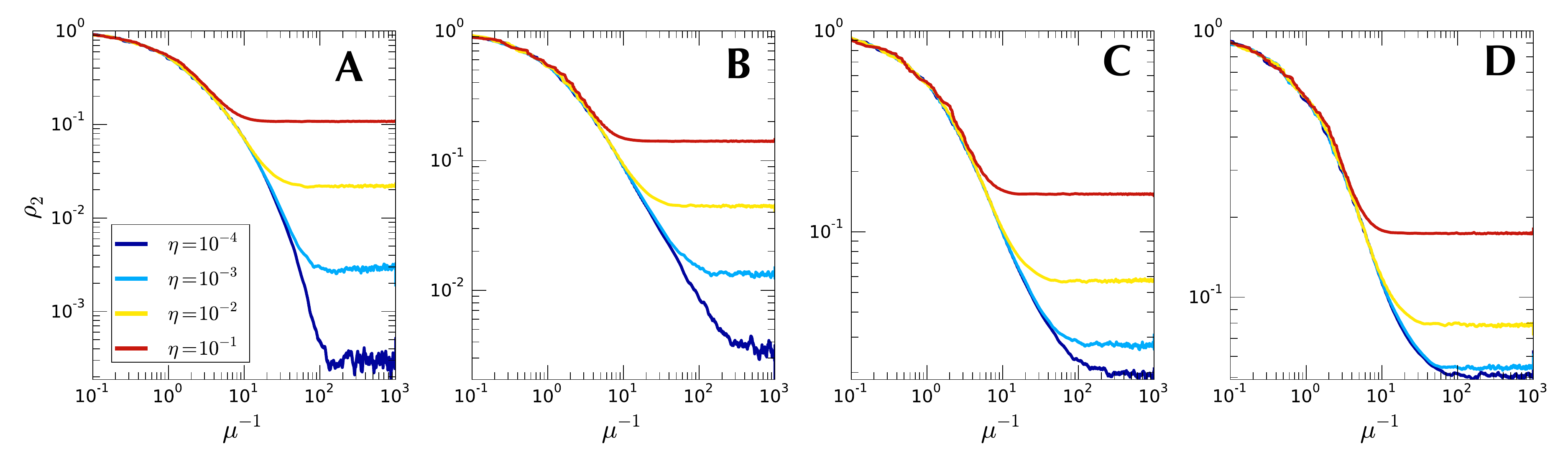}
\end{center}
\caption{Time evolution of the fraction of infected nodes on the second layer for $\mu = 1$, different values of $\eta = 10^{-4}, 10^{-3}, 10^{-2}, 10^{-1}$ and different values of spreading rate: (a) $\lambda = 0.078$, (b) $\lambda=0.083$, (c) $\lambda=0.085$ and (d) $\lambda=0.088$. The multiplex network used is composed of two Erd\"os-R\'enyi networks, both with $n=5\times10^4$, the first layer $\langle k \rangle = 16$ ($(\Lambda_1^1)^{-1} \approx 0.0625 $), while the second $\langle k \rangle = 12$ ($(\Lambda_1^2)^{-1} \approx 0.0833$).}
\label{fig:rho_time}
\end{figure*}

The second peak on the susceptibility curve suggests the existence of a second order phase transition. However, from its existence alone
we cannot conclude this unequivocally, since although this point is related to the delocalization of the disease, the system is already in an endemic phase (upper critical regime in Physics jargon). Observe that if $\eta \in O \left( \frac{1}{N} \right)$, in the thermodynamic limit we would have a phase transition. However, such configuration cannot be considered as a multilayer network, since both layers are (virtually) decoupled. Additionally, observe that we only analyzed layers without correlation. Such features can also introduce different phenomenologies, some were briefly explored in~\cite{deArruda2016}, however for discrete-time.

In order to better understand the second peak of susceptibility we analyze a 2-Layer multiplex network composed by two Erd\H{o}s-R\'enyi networks, in which we can precisely control the mean degree and consequently the epidemic threshold by fixing the number of edges. Furthermore, for scale-free networks with a divergent second moment of its degree distribution, the epidemic threshold vanishes in the thermodynamic limit~\cite{pastor-satorras_epidemic_2015}. On the other hand, Erd\H{o}s-R\'enyi networks always have a non-zero and finite critical point. Aside from that, since the nodes on such a network are statistically equivalent, the probabilities $X_{\beta \tilde{\delta}}$ are expected to be approximately the same. Henceforth we assume that the first layer has a higher connectivity, that is, a lower epidemic threshold.

First of all, analyzing the layers individually for $\frac{\lambda}{\mu} > (\Lambda_1^1)^{-1} \geq \Lambda_1^{-1}$ the first layer is in its upper critical regime (endemic state), while the second layer still is in its sub-critical regime (disease-free state). Then, for a coupling parameter, $\eta > 0$, the probability of a node on the second layer being infected also increases. In fact, for Erd\H{o}s-R\'enyi layers, it will be always larger than zero. Therefore, we can map this problem into an $\epsilon$-SIS model~\cite{Cator2012}, where each node has a probability of experiencing a spontaneous infection. Note that such a model does not present an absorbing state. In this mapping, we are interested on the behavior of the second layer and consider that the self-infection $\epsilon$ is determined by the contribution of the first layer by means of the contacts between nodes in different layers, which are Poisson processes with parameter $\eta$. This would imply that we would not have a second order phase transition. However, we have a transition from a localized system, in which only the first layer is active and able to sustain the disease for long times, to a delocalized system, where both layers are active.

In order to explore the time evolution of the system for a set of parameters near the second susceptibility peak, we run the continuous simulation 50 times and perform a moving average filter over a sampling of the original time series, resulting in $5\times10^4$ points. This approach give us an average curve over time. Note that for continuous simulations the number of points can vary from one run to another. Both networks used have $n=5\times10^4$, the first $\langle k \rangle = 16$ ($(\Lambda_1^1)^{-1} \approx 0.0625 $), while the second $\langle k \rangle = 12$ ($(\Lambda_1^2)^{-1} \approx 0.0833$).

Figure~\ref{fig:rho_time} shows the time evolution of a disease spreading on the second layer for different values of $\lambda$ and $\eta$. The initial conditions for these experiments consider that the first layer has an initial probability of a node being infected equal to $0.01$, while on the second every node is a spreader. Note that we chose this initial condition for visual purposes, since any initial condition would result in a similar steady state regime. In this way, during the transient state we observe a decay of the fraction of infected individuals, then, at the meta state that configures the steady state, we observe a stochastic variation centered on the average value. Besides, such fluctuations tend to increase near a ``critical point''. We observe that for $(\Lambda_1^1)^{-1} > \frac{\lambda}{\mu} > (\Lambda_2^1)^{-1}$ for $\eta = 10^{-4}$ the incidence is very low, of order $O\left(\frac{1}{N}\right)$, however, larger than zero. As we increase the value of $\lambda$ we drive the system to its active state, being able to sustain the disease and spreading it by the intra-edges contacts. Besides, increasing $\eta$ we are able to increase the incidence of the disease due to the intra-edge contacts. Near the critical  point of the second layer, $\frac{\lambda}{\mu} = (\Lambda_2^1)^{-1} = 0.833$, we can observe some features that are similar to a transition. From below, we observe that the lower the value of $\eta$, the longer it takes for the system to reach the steady state, similarly to what it is expected in phase transitions. On the other hand, slightly above the critical point, the time to get into the steady state decreases and the curves for $\eta = 10^{-4}$ and $\eta = 10^{-3}$ get closer. This suggests that the effects of intra-layer spreadings are the main source of spreading. Finally, for $\frac{\lambda}{\mu}$ sufficiently large, we observe the same behavior for all values of $\eta$, i.e. all of them are in an active state.

In addition to the analysis shown in this section, we also inspected in detail the steady state for different system sizes, showing that neither the fluctuations diverge nor the final fraction of infected individuals goes to zero on the second layer. This analysis suggests that we do not have a second order phase transition but that the dynamics changes from a localized to a delocalized phase. In this reach phenomenological scenario, the transition point is still of great importance for practical purposes, for instance when it comes to study immunization policies. These complementary results are shown in Appendix~\ref{sec:app_fss}.

% 3L
\begin{figure}[!t]
\begin{center}
\includegraphics[width=1\linewidth]{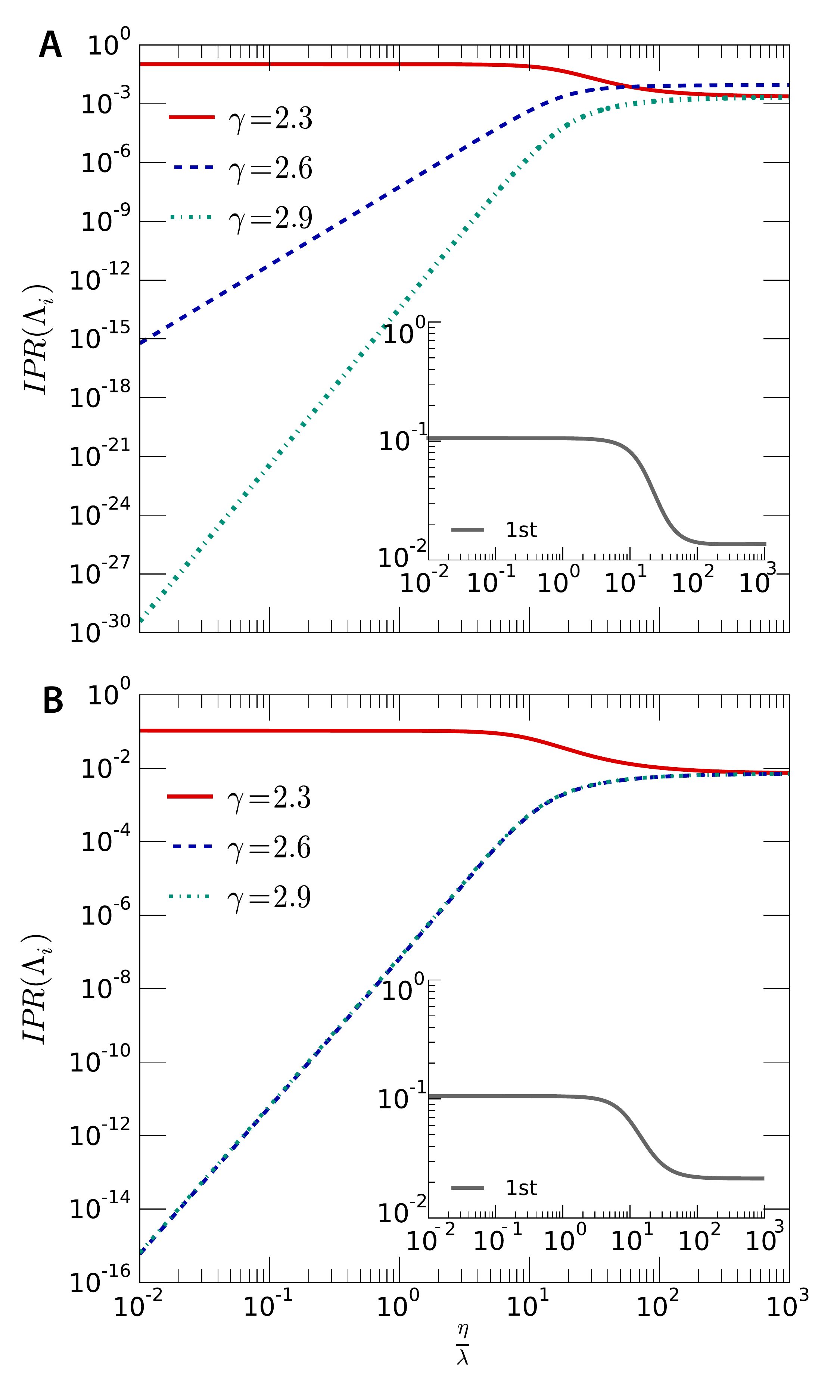}
\end{center}
\caption{Spectral properties of the tensor $\mathcal{R}(\lambda, \eta)$ as a function of the ratio $\frac{\eta}{\lambda}$ for a multiplex with two layers with the same degree distribution (different random realizations of the configuration model) and connected to its counterpart on the other layer. On the top panel we present the inverse participation ratio ($\text{IPR}(\Lambda)$) of the two larger eigenvalues and the individual layer contributions, while on bottom panel we show the leading eigenvalues. Every curve is composed by $10^3$ log spaced points, in order to have enough resolution. On (a) we have the line $(2.3+2.9+2.6)$, while on (b) the multiplex case.}
\label{fig:IPR_Triangle_Line_23_29_26_SF}
\end{figure}

\begin{figure}[!t]
\begin{center}
\includegraphics[width=1\linewidth]{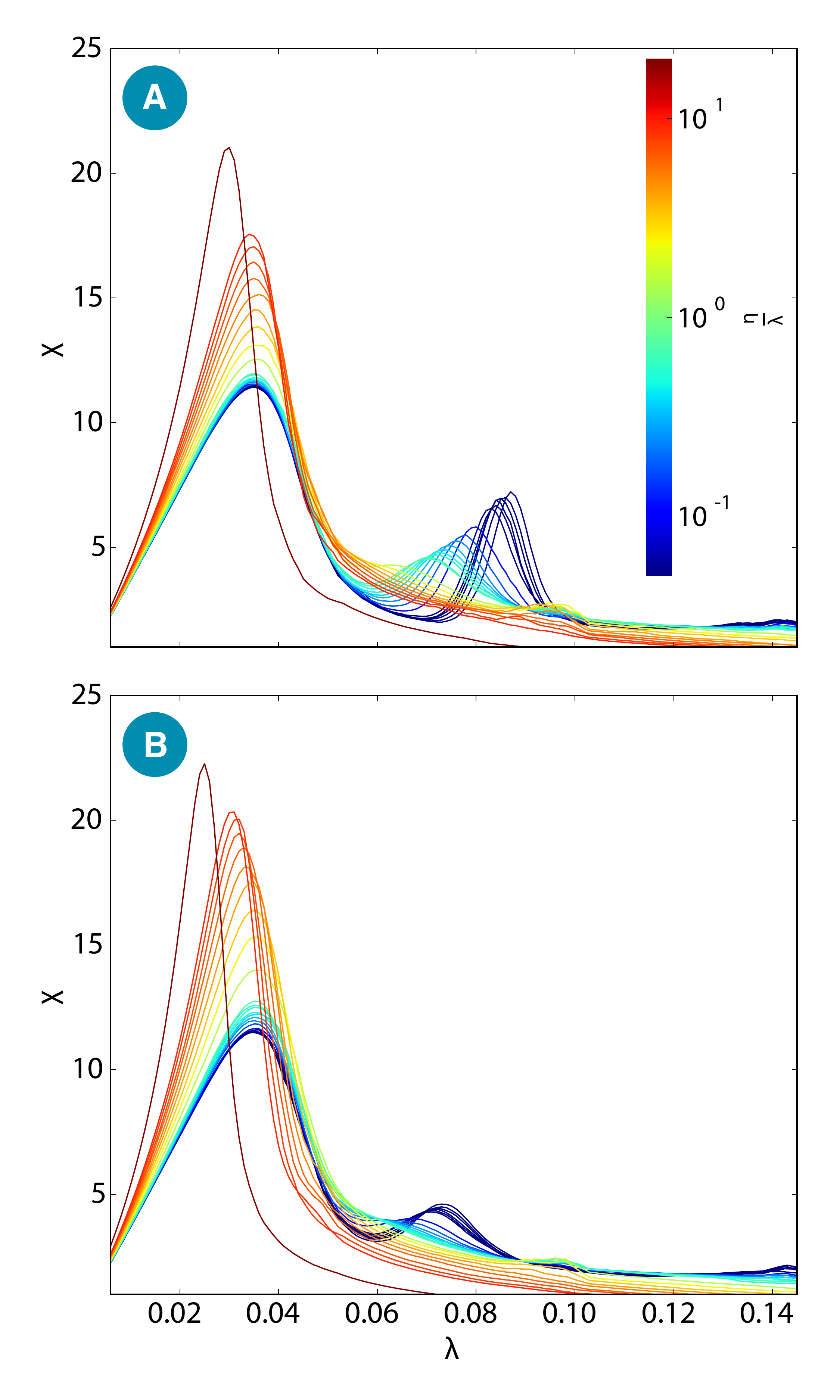}
\end{center}
\caption{Susceptibility $\chi$ as a function of $\lambda$ considering all three layer configurations and many different ratios $\frac{\eta}{\lambda}$, which is represented by the color of the lines. The recovering rate is $\mu=1$. The simulated values are $\frac{\eta}{\lambda} = $ 0.05, 0.06, 0.07, 0.08, 0.09, 0.1, 0.2, 0.3, 0.4, 0.5, 0.6, 0.7, 0.8, 0.9, 1.0, 2, 3, 4, 5, 6, 7, 8, 9, 10, 20. On (a) we have the line $(2.3+2.9+2.6)$, while on (b) the multiplex case.}
\label{fig:X_Triangle_Line_23_29_26_SF}
\end{figure}

\section{3-Layer interconnected systems: the barrier effect} \label{sec:3L}

Following the main ideas of the last sections, we explore the spreading dynamics in multilayer networks with more than two layers. Specifically, we have carried out numerical simulations for a 3-layer system. We generate multiplex networks using three scale-free networks, with $\gamma \approx 2.3$, $\gamma \approx 2.6$ and $\gamma \approx 2.9$, with $\langle k \rangle \approx 8$ and $n = 10^3$ nodes on each layer. Note that we consider three layers with spaced individual leading eigenvalues in order to investigate whether multiple susceptibility peaks are a generic phenomenon of multilayer systems. Note that we have two possible topologies for the network of layers: (i) a line graph and (ii) a triangle (which is a node-aligned multiplex). In its turn, the first can be arranged in three possible configurations by changing the central layer.
That is, we have four possible systems. In this section we focus on two configurations, the multiplex case and the line $(2.3+2.9+2.6)$. Both cases summarize the richness of dynamical processes on interconnected networks, presenting a new phenomenon, the barrier effect of an intermediate layer. We proceed by analyzing the spectral properties of this multilayer system in terms of the inverse participation ratio and the susceptibility. Regarding the other interconnected networks, we present those complementary results and analyses in Appendix~\ref{sec:app_3L}. Additionally, in Appendix~\ref{sec:app_3L_spec} we show that increasing $\frac{\eta}{\lambda}$, also increases the role of the inter-layer edges relative to the intra-layer ones. Consequently, the structure of the network of layers imposes itself more strongly on the eigenvalues of the entire interconnected structure.

\subsection{Spectral analysis} \label{sec:3L_spec}

Figure~\ref{fig:IPR_Triangle_Line_23_29_26_SF} shows the $\text{IPR}(\Lambda_1)$ of tensor $\mathcal{R}$. On the main panel we present the individual contribution of each layer, while on the insets we have the total $\text{IPR}(\Lambda_1)$. On the top panel we have the line $(2.3+2.9+2.6)$, whereas on the bottom panel we have the multiplex network. In this section we focus on the spectral comparison of two cases: (i) the lines $(2.3+2.6+2.9)$ and $(2.3+2.9+2.6)$ and (ii) the line $(2.6+2.3+2.6)$ and the multiplex network. Additionally, the reader is referred to Appendix~\ref{sec:app_3L_spec}, specifically to Fig.~\ref{fig:IPR_Line_23_26_29_SF_Line_26_23_29_SF} for complementary results.

An interesting phenomenon can be observed comparing the different configurations of the network of layers. The largest eigenvalue of the whole system, $\Lambda_1$, has its associated eigenvector localized in the dominant layer, that is, in the layer generated using $\gamma=2.3$. Regarding the line configuration, depending on the position of that layer in the whole system --- i.e., central or peripheral layer --- the contribution of the non-dominant layers to the $\text{IPR}(\Lambda_1)$ varies. In particular, when the dominant layer corresponds to an extreme node of the network of layers, the contribution of the other two layers will be ordered according to the distance to the dominant one. Consequently, when the dominant layer is in the center of the network of layers, the contributions of the non-dominant ones are comparable (see Fig.~\ref{fig:IPR_Line_23_26_29_SF_Line_26_23_29_SF} on Appendix~\ref{sec:app_3L_spec} for complementary results). 

Furthermore, for the first eigenvalue, which is usually enough to analyze the localization as a first order approximation, we observe that the layer with the largest eigenvalue dominates the dynamics. In addition, note the similarities between the multiplex and the line configuration $(2.6+2.3+2.6)$ (see also Fig.~\ref{fig:IPR_Line_23_26_29_SF_Line_26_23_29_SF}, Appendix~\ref{sec:app_3L_spec}), where the non-dominant layers behave similarly. This is because for small values of $\frac{\eta}{\lambda}$, the effect of the extra edge in the network of layers (closing the triangle) is of order $\eta^2$ and so the similar behavior observed for the two configurations. As $\frac{\eta}{\lambda}$ grows, the symmetry in the node-aligned multiplex dominates the eigenvector structure and the contributions of all layers are comparable. As we next show, the different contributions of the layers to the total $\text{IPR}(\Lambda_1)$ are at the root of the multiple susceptibility peaks observed.

\subsection{Multiple susceptibility peaks} \label{sec:3L_multiple}

Figure~\ref{fig:X_Triangle_Line_23_29_26_SF} shows the susceptibility as a function of $\lambda$ for different ratios of $\frac{\eta}{\lambda}$. We observe three well defined peaks on such curves when the ratio $\frac{\eta}{\lambda}$ is small. In addition, similar to the 2-layer case, such peaks tend to become less defined and vanish as the ratio $\frac{\eta}{\lambda}$ increases. The third peak is less defined than the others because the average number of infected nodes is larger in this case. Consequently the susceptibility tend to be lower, since it measures the variance in relation to the average. Such an observation suggests that it could be harder to observe peaks for non-dominating layers that have an individual critical point too far from the dominating layer.

Except for the line $(2.3+2.9+2.6)$ all figures are similar and present similar peaks, implying that the susceptibility peaks occur approximately at the same point (for a complementary analysis see Appendix~\ref{sec:app_X_3L} and Fig.~\ref{fig:X_Line_23_26_29_SF_Line_26_23_29_SF}). On the other hand, the line $(2.3+2.9+2.6)$ shows a slightly different behavior for the second peak, that is found for a larger value of $\lambda$ than for the other cases. This result suggests that when the layer with the largest eigenvalue is located at the center of the line, it can effectively act as a barrier to the disease. In addition, it is verified that the extra inter-edges of the multiplex case does not lead to radical changes on the transition points. We remark that the susceptibility does not measure the fraction of spreaders in the steady state. Thus, despite of the similarities of those curves, the phase diagrams for the incidence of the disease are different. 

Coming back to what is observed for the network of layers described by the line $(2.3+2.9+2.6)$, an interesting phenomenon arises, namely, the formation of barriers to the epidemic spreading. Since the middle layer has the lowest individual eigenvalue among the layers, it creates a barrier effect ``delaying'' the second transition. Moreover, we observe that this transition also vanishes for higher values of the ratio $\frac{\eta}{\lambda}$, if compared to the other cases. This can be related to the inverse participation ratio of $\Lambda_1$, $\text{IPR}(\Lambda_1)$, shown in Fig.~\ref{fig:IPR_Triangle_Line_23_29_26_SF}. Note that, for the line $(2.3+2.9+2.6)$, the contribution of the layer $\gamma = 2.6$ is the lowest. As shown in Section~\ref{sec:2L_numerical} (and in~\cite{Cozzo2013}), for a 2-layer multiplex, the non-dominant layer has its critical point shifted to a lower value of the spreading rate, which means that the outbreak takes place before it would have happened if that layer were isolated. However, here such shifting is compromised by the fact that the central layer is unable to sustain the epidemic process, acting effectively as a barrier for disease contagion. Apart from this new effect, the system behaves qualitatively similar to the 2-layer scenario.

\section{Conclusions}

In this paper, we have generalized and extended previous analyses to the case of multilayer networks. To this end, we have made use of the tensorial representation introduced in~\cite{DeDomenico2013}, which allows to extract upper and lower bounds for the disease incidence of a SIS model and the critical points for both, the SIS and the SIR dynamical processes. We have also validated our analytical insights with extensive numerical simulations, recovering results like those presented in ~\cite{Cozzo2013} regarding the shifting of the global epidemic threshold to lower values of the spreading rate and the role of the so-called dominant layer. Furthermore, we have observed a transition on the spectra of the supra-contact tensor, from the spectra resulting from the union of the individual layers to the spectra of the network of layers. This behavior implies that other dynamics and more complex structures can also be significantly affected by the interconnected nature of the system. In addition, we have also characterized analytically the phenomenon of eigenvalue crossing on the supra-contact tensor for the case of two identical layers. It is worth noticing that any dynamical process that is described by the same matrix will be affected by this effect. 

Our main results concern the emergence and vanishing of multiple susceptibility peaks as a function of the ratio between the inter-layer and intra-layer spreading rates and their relation to the spectral properties of the multilayer, which also revealed the phenomenon of disease localization, and in particular, its relation with the existence of crossings or near-crossings of eigenvalues. Using the QS-Method and Monte Carlo simulations, we have been able to precisely determine the transition points. We remark that the first susceptibility peak is a phase transition, from a disease free state to an endemic, but localized, state. On the other hand, the second peak is a transition from a localized to a delocalized state, which is not a second order phase transition. Additionally, we have proposed an analytical approach based on the use of the inverse participation ratio to characterize such transitions as a localization phenomenon, thus also connecting with \cite{Goltsev2012}. 

A detailed exploration of the parameter space showed that as the ratio between the inter-layer and intra-layer spreading rates increases, the peaks of the susceptibility measured for the non-dominant layers tend to occur at lower values of $\lambda$ and vanish as $\frac{\eta}{\lambda}$ increases up to a point in which only one susceptibility peak is observed, which is a true phase transition. Interesting enough, our results point out that such a transition can take place for even lower values of $\lambda$ than the inverse of the largest leading eigenvalue among all individual layers. 

Finally, another important finding presented here is the opposite phenomenon, namely, the barrier effect, which happens when the susceptibility peak takes place at a larger value of $\lambda$ than that expected as a consequence of the multiplex topology. Specifically, if the layers are arranged in such a way that the one with the smallest leading eigenvalue is at the center of the network of layers (for instance, as it happens for the line $(2.3+2.9+2.6)$ configuration), then the corresponding transition could be delayed due to the barrier effect. Summarizing, our results emphasize the importance of studying multilayer systems as they are and not only as a collection of individual layers.

\acknowledgments
FAR acknowledge CNPq (grant 305940/2010-4), Fapesp (2013/26416-9). GFA acknowledges Fapesp for the sponsorship provided (grants 2012/25219-2 and 2015/07463-1). E. C was supported by the FPI program of the Government of Arag\'on, Spain. Y. M. acknowledges support from the Government of Arag\'on, Spain through a grant to the group FENOL, by MINECO and FEDER funds (grant FIS2014-55867-P) and by the European Commission FET-Proactive Project Multiplex (grant 317532). This research was developed using the computational resources of Centro de Ci\^encias Matem\'aticas Aplicadas \`a Ind\'ustria (CeMEAI) supported by FAPESP.

\appendix

\section{Tensorial representation} \label{sec:app_tensor}

In this appendix we extend some important concepts of the tensorial representation. On Section~\ref{sec:app_tensor_projection} we present the projections, while on Section~\ref{sec:app_eig_sec} we show the equivalence of the eigentensorial problem and the eigenvector problem of the supra adjacency matrix. Finally on Section~\ref{sec:Appendix_A} we prove the relation between the tensorial projection and the matricial representation, which is fundamental to the interlacing results.

\subsection{Tensorial projections} \label{sec:app_tensor_projection}

For the sake of completeness we present other projections of multilayer networks, which are specially convenient on tensorial notation, due to its compactness. Besides the adjacency tensor presented on the main text, the network of layers \cite{Garcia2014} also characterizes the topology of the system. In this reduced network representation, each node represents one layer and the edges between them codify the number of edges connecting those two layers. Formally we have,
\begin{equation}
  \Psi_{\tilde{\delta}}^{\tilde{\gamma}} = M_{\beta \tilde{\delta}}^{\alpha \tilde{\gamma}} U^{\beta}_{\alpha},
\end{equation}
where $\Psi_{\tilde{\delta}}^{\tilde{\gamma}} \in \mathbb{R}^{m \times m}$. Note that such a network presents self-loops, which are weighted by the number of edges on the layer. Additionally, since we assume that the layers have the same number of nodes, the edges of the network of layers have weights equal to the number of nodes $n$.

Another important reduction of the multilayer network is the so-called projection~\cite{DeDomenico2013}. Such network aggregates all the information into one layer, including self-loops that stand for the number of layers in which a node appears. Mathematically, we have
\begin{equation} \label{eq:agregated}
 P_\beta^\alpha = M_{\beta \tilde{\gamma}}^{\alpha \tilde{\delta}} U_{\tilde{\delta}}^{\tilde{\gamma}},
\end{equation}
where $P_\beta^\alpha \in \mathbb{R}^{n \times n}$.

\subsection{Eigenvalue problem} \label{sec:app_eig_sec}

As presented on the main text, the epidemic threshold is closely related to the
leading eigenvalues of the supra-contact tensor. Here we describe the
eigenvalue problem considering the tensorial representation. Such eigenvalue problem can be generalized to the case of a rank-4 tensor leading to
\begin{equation}
 \mathcal{R}_{\beta \tilde{\delta}}^{\alpha \tilde{\gamma}}  f_{\alpha \tilde{\gamma}}(\Lambda) = \Lambda f_{\beta \tilde{\delta}}(\Lambda),
\end{equation}
where $\Lambda$ is an eigenvalue and $f_{\beta \tilde{\delta}}(\Lambda)$
is the corresponding eigentensor. In addition, we are assuming that the
eigentensors form an orthonormal basis. Importantly, the supra-contact
matrix, $R$, in ~\cite{Cozzo2013} can be understood as a
flattened version of the tensor $\mathcal{R}_{\beta
\tilde{\delta}}^{\alpha \tilde{\gamma}}(\lambda, \eta)$. Consequently,
all the results for $R$ also apply to the tensor $\mathcal{R}$. As
argued in~\cite{DeDomenico2013}, that supra-adjacency matrix
corresponds to unique unfolding of the fourth-order tensor $m$ yielding square matrices. Following this unique mapping we have the correspondence of the eigensystems. Here, we consider that the
eigenvalues are ordered as $\Lambda_1 \geq \Lambda_2 \geq
... \Lambda_{nm}$ and denote the individual layer eigenvalues as
$\Lambda^l_{i}$.

\subsection{Proof of Equation~\ref{eq:Net_net}}\label{sec:Appendix_A}

Considering the matricial representation of a multilayer network, given by
\begin{equation}
 A = \oplus_{\alpha} A^\alpha + C = 
 \begin{bmatrix}
       A_1 & C_{12} & \cdots & C_{1m} \\
       C_{21} & A_2 & \cdots & C_{2m} \\
      \vdots & \vdots & \ddots & \vdots \\
       C_{m1} & C_{m2} & \cdots & A_m \\
     \end{bmatrix}
\end{equation}
where $A \in \mathbb{R}^{nm \times nm}$, $A^\alpha \in \mathbb{R}^{n \times n}$ is the adjacency matrix of the layer $\alpha \in \{ 1, 2, ... m\}$ and $C$ is a coupling matrix. Since we assume multilayer network in which the layers have the same number of nodes we have $C_{ij} = I$. Assuming a partition of such network, represented by $S \in \mathbb{R}^{nm \times m}$, which is the characteristic matrix of such partition, where $S_{ij} = 1$ if $i \in V_j$ and zero otherwise, where $V_j$ is the network of layers partition.

In order to use the results of~\cite{Garcia2014, Cozzo2015} we have to prove that the network of layers matrix $\bar{R}$~\cite{Garcia2014, Cozzo2015} is an unfolding of our tensor $\Phi_{\tilde{\delta}}^{\tilde{\gamma}}(\lambda, \eta)$, formally given by
\begin{equation}
 \bar{R} = \Gamma^{-1} S^T A S,
\end{equation}
where $\Gamma$ is a diagonal matrix with normalizing constants (for more, see references~\cite{Garcia2014, Cozzo2015}). In words, the product $A S$ is a summation over the blocks of the matrix $A$, resulting in a matrix with the degree of each node. The subsequent left product with $S^T$ impose another summation, whose result is a matrix composed by the sum of all elements of the blocks. Finally, the product by $\Gamma^{-1}$ normalize the result by $\frac{1}{n}$. Formally we have,
\begin{equation}
AS =  \begin{bmatrix}
       k^{11} & k^{12} & \cdots & k^{1m} \\
       k^{21} & k^{22} & \cdots & k^{2m} \\
      \vdots & \vdots & \ddots & \vdots \\
       k^{m1} & k^{m2} & \cdots & k^{mm},
     \end{bmatrix}
\end{equation}
where $k^{ij} \in \mathbb{R}^{n \times 1}$ is a vector with the number of edges emanating from each node on layer $i$ to layer $j$ and $AS \in \mathbb{R}^{nm \times m}$. Then,
\begin{equation}
S^TAS =  \begin{bmatrix}
       \sum k^{11} & \sum k^{12} & \cdots & \sum k^{1m} \\
       \sum k^{21} & \sum k^{22} & \cdots & \sum k^{2m} \\
      \vdots & \vdots & \ddots & \vdots \\
       \sum k^{m1} & \sum k^{m2} & \cdots & \sum k^{mm},
     \end{bmatrix}
\end{equation}
where $\sum k^{ij} \in \mathbb{R}$ are scalars with the number of edges that connect a node on layer $i$ to a node on layer $j$. Finally, the product by $\Gamma^{-1}$ introduce the average degree instead of the summation, producing the same results as Eq.~\ref{eq:Net_net}.

\section{The Susceptible-Infected-Susceptible (SIS) model analysis} \label{sec:app_sis}

In this section we present an extension of the analysis presented on the main text regarding the SIS model. To begin with, we present comments on the exact model definition and its relation with the first order approximation on Section~\ref{sec:app_sis_comp}. On Section~\ref{sec:app_sis_thr} we present a derivation of the critical point for the first order approximation, while on Section~\ref{sec:app_sis_bounds} we present the derivation of the lower and upper bound for such model.

\subsection{Model definition: complementary comments} \label{sec:app_sis_comp}

In probability theory and stochastic processes it is usual to define random variables as capital letter. However, that is the same usual notation for tensors. In order to avoid confusion we will use bold capital letters for random variables. For instance, we define the Bernoulli random variable that defines the state of a node as $\boldsymbol{S_{\beta \tilde{\delta}}}$, where it assumes one of two values, zero if the node $\beta \tilde{\delta}$ is susceptible or one if it is infected. By definition, $X_{\beta \tilde{\delta}} = \langle \boldsymbol{S_{\beta \tilde{\delta}}} \rangle$, where $\langle \cdot \rangle$ is the expectation operator and $X_{\beta \tilde{\delta}}$ is the probability of the node $\beta \tilde{\delta}$ being infected.

In this way, without any assumption on the independence of random variables the exact equation can be written as 
\begin{equation} \label{eq:SIS_tensor_exact}
\dfrac{d \langle \boldsymbol{S_{\beta \tilde{\delta}}} \rangle}{dt} = \la - \mu \boldsymbol{S_{\beta \tilde{\delta}}} +\left( 1 - \boldsymbol{S_{\beta \tilde{\delta}}} \right) \lambda \mathcal{R}_{\beta \tilde{\delta}}^{\alpha \tilde{\gamma}}(\lambda, \eta)  \boldsymbol{S_{\alpha \tilde{\gamma}}} \ra,
\end{equation}
where the supra contact tensor is defined in~\ref{eq:adj_tensor}. This equation can be interpreted as an exact version of the epidemic process~\cite{Mieghem09}. However without any approximation the solution of such problem involves $O(2^{nm})$ equations, since we have to write the expressions for the expectation for all the products. The first order approximation consists in $\langle \boldsymbol{S_{\beta \tilde{\delta}}} \boldsymbol{S_{\alpha \tilde{\gamma}}} \rangle \approx \langle \boldsymbol{S_{\beta \tilde{\delta}}} \rangle \langle \boldsymbol{S_{\alpha \tilde{\gamma}}} \rangle = X_{\beta \tilde{\delta}} X_{\alpha \tilde{\gamma}}$. Such approximation is shown on eq.~\ref{eq:SIS_tensor}. Interestingly, observe that eq.~\ref{eq:SIS_tensor_exact} can be written in terms of the covariance, defined as $\text{Cov}[\boldsymbol{S_{\beta \tilde{\delta}}}, \boldsymbol{S_{\alpha \tilde{\gamma}}}] = \langle \boldsymbol{S_{\beta \tilde{\delta}}} \boldsymbol{S_{\alpha \tilde{\gamma}}} \rangle - \langle \boldsymbol{S_{\beta \tilde{\delta}}} \rangle \langle \boldsymbol{S_{\alpha \tilde{\gamma}}} \rangle$. Consequently, isolating the probability of the product and substituting it in eq.~\ref{eq:SIS_tensor_exact} we find, by inspection, that the error is given by $\text{Cov}[\boldsymbol{S_{\beta \tilde{\delta}}}, \boldsymbol{S_{\alpha \tilde{\gamma}}}]$, which is assumed to be zero. In~\cite{Bovenkamp2015}, the authors observed this relation and proposed an accuracy criteria for monoplex networks.

\subsection{The epidemic threshold} \label{sec:app_sis_thr}

An important concept for dynamical systems that present an absorbing state and an active phase is the critical point. Considering the SIS process, below this point the system is inactive and the disease tends to disappear. On the other hand, for above this point we have the active phase, where the disease is present on a fraction of the population. Assuming $\mu > 0$ and that the dynamics has reached the steady state,
$\dfrac{d X_{\beta \tilde{\delta}}}{dt} = 0$, we can write
eq.~\ref{eq:SIS_tensor} as
\begin{equation}
 \frac{X_{\beta \tilde{\delta}}^{\infty}}{ 1 - X_{\beta \tilde{\delta}}^{\infty}} = \left(\frac{\lambda}{\mu} \right) \mathcal{R}_{\beta \tilde{\delta}}^{\alpha \tilde{\gamma}} X_{\alpha \tilde{\gamma}}^{\infty}.
\end{equation}
Expanding the left-hand term following the geometrical series, where $\frac{X_{\beta \tilde{\delta}}^{\infty}}{ 1 - X_{\beta \tilde{\delta}}^{\infty}} = \sum_{k=1}^{\infty} \left( X_{\beta \tilde{\delta}}^{\infty} \right)^k$ for $X_{\beta \tilde{\delta}}^{\infty} < 1$, we obtain 
\begin{equation} \label{eq:sis_serie}
\left(\frac{\mu}{\lambda} \right) \sum_{k=1}^{\infty} \left( X_{\beta \tilde{\delta}}^{\infty} \right)^k = \mathcal{R}_{\beta \tilde{\delta}}^{\alpha \tilde{\gamma}}(\lambda, \eta)  X_{\alpha \tilde{\gamma}}^{\infty}.
\end{equation}

In addition, similarly to~\cite{Mieghem09}, suppose $X_{\beta \tilde{\delta}}^{\infty} = \epsilon f_{\beta \tilde{\delta}}$, where $\epsilon$ is an arbitrary small constant and $f_{\beta \tilde{\delta}} \geq 0$. Substituting in eq.~\ref{eq:sis_serie} and dividing by $\epsilon$ we have
\begin{equation}
\mathcal{R}_{\beta \tilde{\delta}}^{\alpha \tilde{\gamma}}(\lambda, \eta)  f_{\alpha \tilde{\gamma}} = \left(\frac{\mu}{\lambda} \right) f_{\beta \tilde{\delta}} + \epsilon \left(\frac{\mu}{\lambda} \right) \left( f_{\beta \tilde{\delta}} \right)^2 + \mathcal{O}(\epsilon^2).
\end{equation}
Considering a sufficiently small $\epsilon > 0$ this expression reduces to the eigentensor equation
\begin{equation}
\mathcal{R}_{\beta \tilde{\delta}}^{\alpha \tilde{\gamma}}(\lambda, \eta)  f_{\alpha \tilde{\gamma}} = \left(\frac{\mu}{\lambda} \right) f_{\beta \tilde{\delta}},
\end{equation}
leading to the critical point
\begin{equation} \label{eq:threshold2}
 \left(\frac{\mu}{\lambda} \right)_c = \Lambda_1.
\end{equation}
where $\Lambda_1$ is the largest eigenvalue of $\mathcal{R}$, which is the same as the largest eigenvalue of $R$ in~\cite{Cozzo2013}.

\subsection{Upper and lower bounds for the steady-state} \label{sec:app_sis_bounds}

In order to obtain some bounds for the epidemic incidence considering the steady state, where $\dfrac{d X_{\beta \tilde{\delta}}}{dt} = 0$. For a monolayer system those bounds were calculated in~\cite{Mieghem09}. We consider a multilayer network without self loops and denote the steady state of each node as $X_{\beta \tilde{\delta}}^{\infty}$. Then, imposing $\dfrac{d X_{\beta \tilde{\delta}}}{dt} = 0$ to Eq.~\ref{eq:SIS_tensor}  we have
\begin{equation} \label{eq:steady}
\begin{split}
  X_{\beta \tilde{\delta}}^{\infty} = \frac{\lambda \mathcal{R}_{\beta \tilde{\delta}}^{\alpha \tilde{\gamma}}(\lambda, \eta)  X^{\infty}_{\alpha \tilde{\gamma}}}{\lambda \mathcal{R}_{\beta \tilde{\delta}}^{\alpha \tilde{\gamma}}(\lambda, \eta)  X^{\infty}_{\alpha \tilde{\gamma}} + \mu}    = 1 - \frac{1}{\frac{\lambda}{\mu} \mathcal{R}_{\beta \tilde{\delta}}^{\alpha \tilde{\gamma}}(\lambda, \eta)  X^{\infty}_{\alpha \tilde{\gamma}} + 1}.
\end{split}
\end{equation}
The value of $X_{\beta \tilde{\delta}}^{\infty}$ is then obtained by
iterating the above equation from an initial value, until convergence.
Upper and lower bounds can be obtained by considering only the first
iteration of Eq.~\ref{eq:steady}. For the upper bound we have
\begin{equation} \label{eq:upper}
 X_{\beta \tilde{\delta}}^{\infty} \leq 1 - \frac{1}{ \left( \frac{\lambda}{\mu} \right) d_{\beta \tilde{\delta}} + 1}.
\end{equation}
where
\begin{equation} \label{eq:d_beta_delta}
\begin{split}
  d_{\beta \tilde{\delta}}& = \mathcal{R}_{\beta \tilde{\delta}}^{\alpha \tilde{\gamma}}(\lambda, \eta)  U_{\alpha \tilde{\gamma}} = \\
 & = M_{\beta \tilde{\gamma}}^{\alpha \tilde{\xi}} E_{\tilde{\xi}}^{\tilde{\gamma}}(\tilde{\delta}\tilde{\delta}) u_{\alpha} + \frac{\eta}{\lambda} M_{\nu \tilde{\delta}}^{\xi \tilde{\gamma}} E_{\xi}^{\nu}(\beta \beta) u_{\tilde{\gamma}}.
\end{split}
\end{equation}
As can be noticed, there are two different contributions to the upper
bound coming from intra and inter-layers connectivity. Both of them tend to increase
the probability of a node being infected. Furthermore, the higher is the
degree, the higher is this upper bound. On the other hand, for the lower
bound, let us denote $\text{Min}\{X_{\beta \tilde{\delta}}^{\infty}\}
=X^{\text{min}}$. Then, substituting $X^{\text{min}}$ in
Eq.~\ref{eq:steady} we have
\begin{equation} 
\begin{split}
 X^{\text{min}} \geq 1 -\frac{1}{\frac{\lambda}{\mu} \mathcal{R}_{\beta \tilde{\delta}}^{\alpha \tilde{\gamma}}(\lambda, \eta)  U_{\alpha \tilde{\gamma}} X^{\text{min}} + 1}.
\end{split}
\end{equation}
Denoting $\text{Min}\{d_{\beta \tilde{\delta}}\} = d^{\text{min}}$, we obtain
\begin{equation}
 X^{\text{min}} \geq 1 - \frac{1}{ \left( \frac{\lambda}{\mu} \right) d^{\text{min}}},
\end{equation}
which can be inserted into Eq.~\ref{eq:steady} to give,
\begin{equation} \label{eq:lower}
 X^{\infty}_{\beta \tilde{\delta}} \geq X^{\text{min}} \geq 1 - \frac{1}{1 + \frac{d_{\beta \tilde{\delta}}}{d^{\text{min}}} \left[ \left( \frac{\lambda}{\mu} \right)  d^{\text{min}} - 1 \right]}.
\end{equation}

Finally, combining Eqs.~\ref{eq:upper} and~\ref{eq:lower}, the bounds of Eq.~\ref{eq:SIS_tensor} are
\begin{equation} \label{eq:bounds2}
 1 - \frac{1}{1 + \frac{d_{\beta \tilde{\delta}}}{d^{\text{min}}} \left[ \left( \frac{\lambda}{\mu} \right) d^{\text{min}} - 1 \right]} \leq X^{\infty}_{\beta \tilde{\delta}} \leq 1 - \frac{1}{\left( \frac{\lambda}{\mu} \right) d_{\beta \tilde{\delta}} + 1}.
\end{equation}

\section{The Susceptible-Infected-Recovered (SIR) Model} \label{sec:app_sir}

Aside from the SIS epidemic model, we can also consider the SIR model. Contrasting with the SIS, which have just one absorbing state (inactive), the SIR have many absorbing states. In fact, considering an infinite population we have an infinite number of absorbing states.

\subsection{Model definition}

Introducing the recovered and susceptible states, here denoted by $Y_{\beta \tilde{\delta}}$ and $Z_{\beta \tilde{\delta}}$, respectively. Then, using a similar notation as in the latter section and associating Poisson processes to nodes and edges, we have the  dynamical set of equations
\begin{align} \label{eq:SIR_tensor}
  \begin{split}
 \dfrac{d X_{\beta \tilde{\delta}}}{dt} &= -\mu X_{\beta \tilde{\delta}} + Z_{\beta \tilde{\delta}} \lambda \mathcal{R}_{\beta \tilde{\delta}}^{\alpha \tilde{\gamma}}(\lambda, \eta)  X_{\alpha \tilde{\gamma}} \\ 
 \dfrac{d Y_{\beta \tilde{\delta}}}{dt} &= \mu X_{\beta \tilde{\delta}} \\
 \dfrac{d Z_{\beta \tilde{\delta}}}{dt} &= - Z_{\beta \tilde{\delta}} \lambda \mathcal{R}_{\beta \tilde{\delta}}^{\alpha \tilde{\gamma}}(\lambda, \eta)  X_{\alpha \tilde{\gamma}}.
  \end{split}
\end{align}
Note that the Poisson processes on the nodes model the recovering, whereas on the edges, model the spreading.

\subsection{Epidemic threshold}

Since there is no dynamic steady state in the SIR model, the epidemic threshold has a different interpretation from that of the SIS model. Above the threshold the total number of recovered individuals reaches a finite fraction of the population, when the dynamic starts with a small fraction of infected individuals. Formally, the initial condition are: $X_{\beta \tilde{\delta}}(0) = \frac{c}{nm}$, $Y_{\beta \tilde{\delta}}(0) = 0$ and $Z_{\beta \tilde{\delta}}(0) = 1 - \frac{c}{nm}$, where $c$ is a small constant, $c \ll nm$. Neglecting higher order terms, we have 
\begin{equation}
 \dfrac{d X_{\beta \tilde{\delta}}}{dt} = -\mu X_{\beta \tilde{\delta}} + \lambda \mathcal{R}_{\beta \tilde{\delta}}^{\alpha \tilde{\gamma}}(\lambda, \eta)  X_{\alpha \tilde{\gamma}}.
\end{equation}
After a proper factorization, 
\begin{equation}
 \dfrac{d X_{\beta \tilde{\delta}}}{dt} =  \lambda \left( \mathcal{R}_{\beta \tilde{\delta}}^{\alpha \tilde{\gamma}}(\lambda, \eta) - \frac{\mu}{\lambda} \delta_{\beta \tilde{\delta}}^{\alpha \tilde{\gamma}} \right) X_{\alpha \tilde{\gamma}},
\end{equation}
where $\delta_{\beta \tilde{\delta}}^{\alpha \tilde{\gamma}}$ is a tensor analogous to the identity matrix, whose elements are one if the indices are the same. The epidemic threshold is as in eq.~\ref{eq:threshold}, which is the critical value for both SIR and SIS dynamics.

\section{2-Layer Multiplex systems} \label{sec:app_2L}

In this section we present some complementary analysis for the 2-Layer multiplex case. Here we focus on some spectral aspects of such systems, mainly on the eigenvalue crossing and near-crossing phenomenon, presented on Section~\ref{sec:2L_spectral} and additionally on the second susceptibility peak using a finite size analysis. Such results are presented on Section~\ref{sec:app_fss} and are complementary to Section~\ref{sec:sec_peak} on the main text.

\subsection{Spectral aspects} \label{sec:2L_spectral}

In this section we focus on the spectral analysis of the tensor $\mathcal{R}(\lambda, \eta)$ as a function of the ratio $\frac{\eta}{\lambda}$. First of all, we present an analytical approach to the problem of eigenvalue crossings on Section~\ref{sec:app_crossing}, then we focus on three special cases in increasing order of complexity: (i) the identical case, where both layers are exactly the same. Thus, there is a high correlation between the degree on each layer, presented on Section~\ref{sec:app_identical}; (ii) the non-identical case, where both layers present the same degree distribution, but different configurations on Section~\ref{sec:app_similar}. The case of two different layer structures, considering that their leading eigenvalues are spaced was presented on the main text.

% Spectra
\begin{figure}[!t]
\begin{center}
\includegraphics[width=1\linewidth]{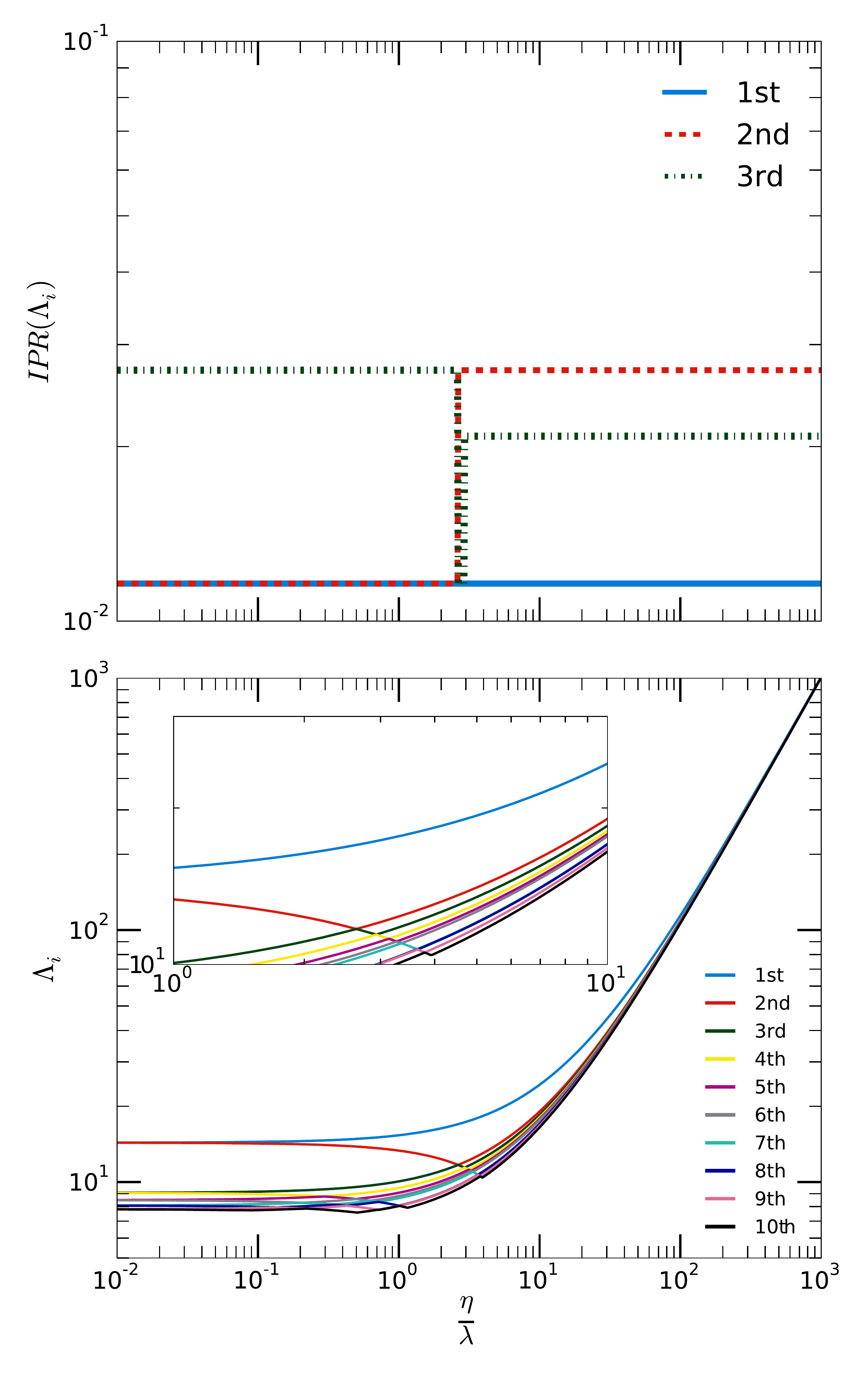}
\end{center}
\caption{Spectral properties of the tensor $\mathcal{R}(\lambda, \eta)$ as a function of the ratio $\frac{\eta}{\lambda}$ for a multiplex with two layers with the exact same degree distribution and connected to its counterpart on the other layer. On the top panel we present the inverse participation ratio ($\text{IPR}(\Lambda)$) of the three larger eigenvalues, while on bottom panel we show the leading eigenvalues. Every curve is composed by $10^3$ log spaced points, in order to have enough resolution.}
\label{fig:Localization_Identical}
\end{figure}

\begin{figure}[!t]
\begin{center}
\includegraphics[width=1\linewidth]{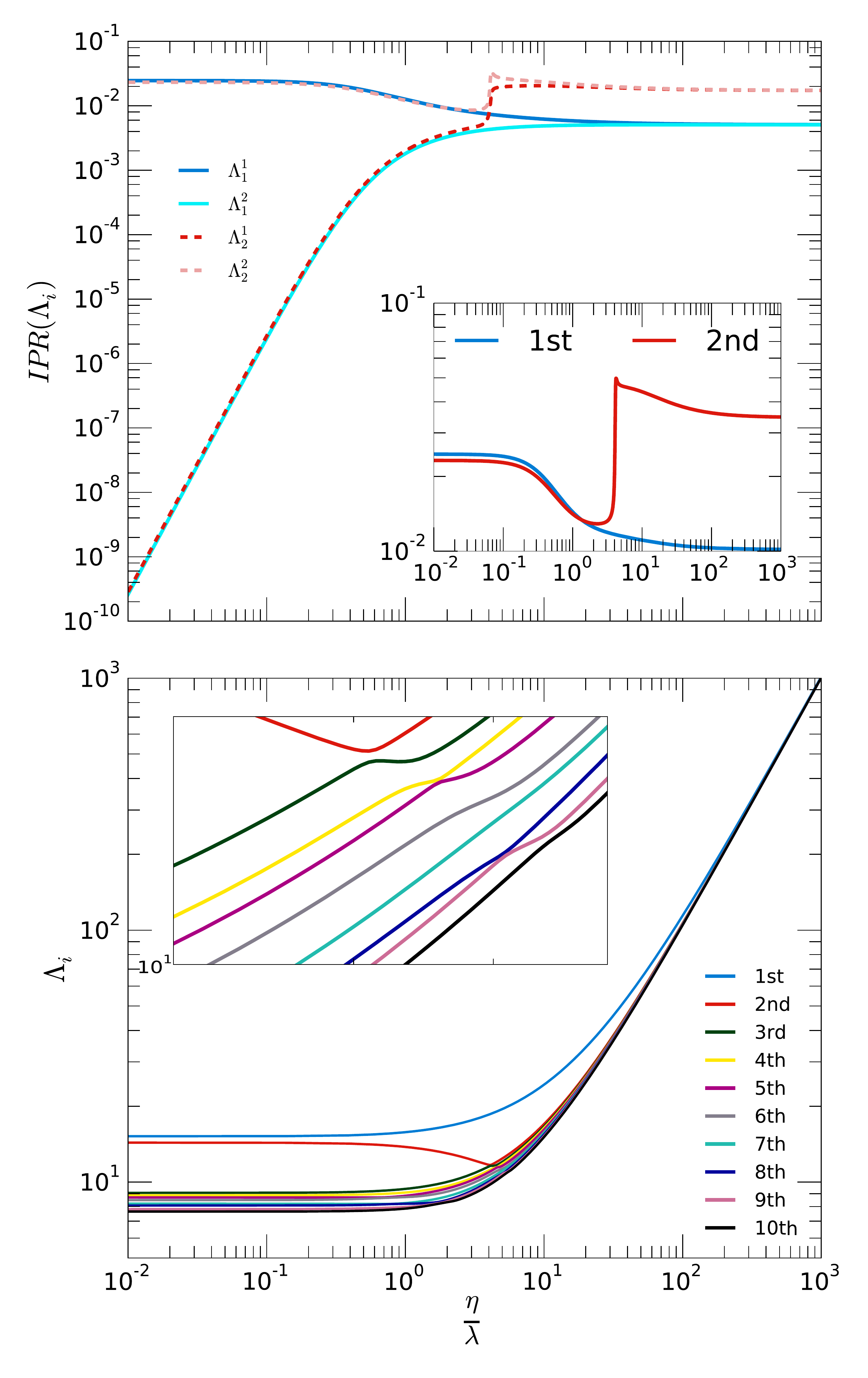}
\end{center}
\caption{Spectral properties of the tensor $\mathcal{R}(\lambda, \eta)$ as a function of the ratio $\frac{\eta}{\lambda}$ for a multiplex with two layers with the same degree distribution (different random realizations of the configuration model) and connected to its counterpart on the other layer. On the top panel we present the inverse participation ratio ($\text{IPR}(\Lambda)$) of the two larger eigenvalues and the individual layer contributions, while on bottom panel we show the leading eigenvalues. Every curve is composed by $10^3$ log spaced points, in order to have enough resolution.}
\label{fig:Localization_Similar}
\end{figure}

\subsubsection{Eigenvalue crossing} \label{sec:app_crossing}

Let us analyze the spectra of a simple setup: multiplex networks
composed by $l$ identical layers. Such class of networks provides
insights about the spectral behavior as a function of $\left(
\frac{\eta}{\lambda} \right)$. Although they are not very realistic a
priori, there are situations in which this representation is helpful:
for instance, in the context of disease contagion, one might think of a
multi-strain disease in which each strain propagates in a different
layer allowing co-infection of the host population.

The adjacency tensor can be written as
\begin{equation}
 \mathcal{R}_{\beta \tilde{\delta}}^{\alpha \tilde{\gamma}}(\lambda, \eta) = A_{\beta}^{\alpha} \delta^{\tilde{\gamma}}_{\tilde{\delta}} + \frac{\eta}{\lambda} \delta_{\beta}^{\alpha} K^{\tilde{\gamma}}_{\tilde{\delta}},
\end{equation}
where $A_{\beta}^{\alpha}$ is the 2-rank layer adjacency tensor, $K_{\tilde{\gamma}}^{\tilde{\delta}}$ is the adjacency tensor of the network of layers, which is a complete graph on the multiplex case, and $\delta_{\beta}^{\alpha}$ is the Kronecker delta. Observe that  the sum of two Kronecker products, $\bar{A} = I_m \otimes A + \frac{\eta}{\lambda} K_m \otimes I_n$, where $I_n$ is the identity matrix of size $n$ and $K_m$ is the adjacency matrix of the complete graph with $m$ nodes is the unfolding of the adjacency tensor in this case. In this way, the eigenvalue problem can be written as 
\begin{equation}
  \mathcal{R}_{\beta \tilde{\delta}}^{\alpha \tilde{\gamma}} f_{\alpha \tilde{\gamma}} = A_{\beta}^{\alpha} \delta^{\tilde{\gamma}}_{\tilde{\delta}} f_{\alpha \tilde{\gamma}} + \frac{\eta}{\lambda} \delta_{\beta}^{\alpha} K^{\tilde{\gamma}}_{\tilde{\delta}} f_{\alpha \tilde{\gamma}},
\end{equation}
where the sum of the eigenvalues of $A$, $\Lambda^l_i$, and $K$, $\mu_i$, are also eigenvalues of the adjacency tensor, hence $\mathcal{R}_{\beta \tilde{\delta}}^{\alpha \tilde{\gamma}} f_{\alpha \tilde{\gamma}} = \left( \Lambda^l_i +  \frac{\eta}{\lambda} \mu_j \right) f_{\alpha \tilde{\gamma}}$, $i=1,2,...n$ and $j = 1,2,...m$. Then, 
\begin{equation}
  \left( \Lambda^l_i +  \frac{\eta}{\lambda} \mu_j \right)  = \left( \Lambda^l_k +  \frac{\eta}{\lambda} \mu_s \right).
\end{equation}

The eigenvalues of the complete graph are $\mu_1 = m-1$, and $\mu_i = -1, \hspace{0.1cm} \forall i > 1$, yielding to
\begin{equation}
 \frac{\eta}{\lambda} = \frac{\Lambda^l_k - \Lambda^l_i}{m},
\end{equation}
which imposes crossings on the eigenvalues of the adjacency tensor for identical layers, since $\left( \frac{\eta}{\lambda} \right)$ is a continuous parameter.

% FSS
\begin{figure*}[!ht]
\begin{center}
\includegraphics[width=1\linewidth]{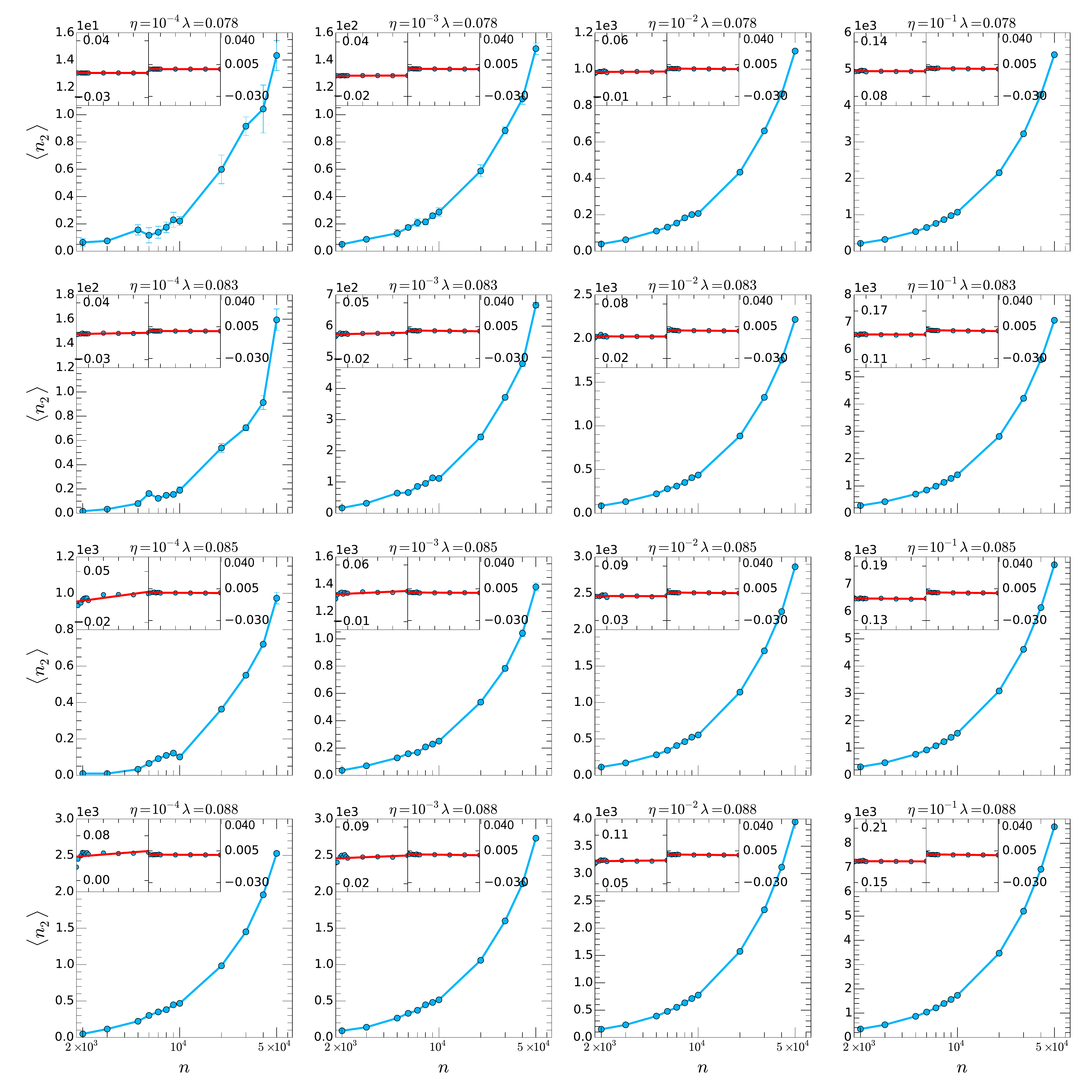}
\end{center}
\caption{Final number of infected nodes on the second layer (with lowest individual eigenvalue) as a function of the size of the layers on the main panels, while on the insets we present the fraction of infected nodes on the left and the standard deviation on the steady state on the right. The parameters used on the simulations are shown on the tile of each panel. They are a combination of the parameters $\lambda = 0.078, 0.083, 0.085, 0.088$ and $\eta = 10^{-4}, 10^{-3}, 10^{-2}, 10^{-1}$. Furthermore, the layer sizes are $n = 2 \times 10^{3}, 3 \times 10^{3}, 4 \times 10^{3}, 5 \times 10^{3}, 6 \times 10^{3}, 7 \times 10^{3}, 8 \times 10^{3}, 9 \times 10^{3}, 10^{4}, 2 \times 10^{4}, 3 \times 10^{4}, 4 \times 10^{4}$ and $5 \times 10^{4}$ and $m = 2$ on all cases.}
\label{fig:FSS}
\end{figure*}

\begin{figure}[!t]
\begin{center}
\includegraphics[width=1\linewidth]{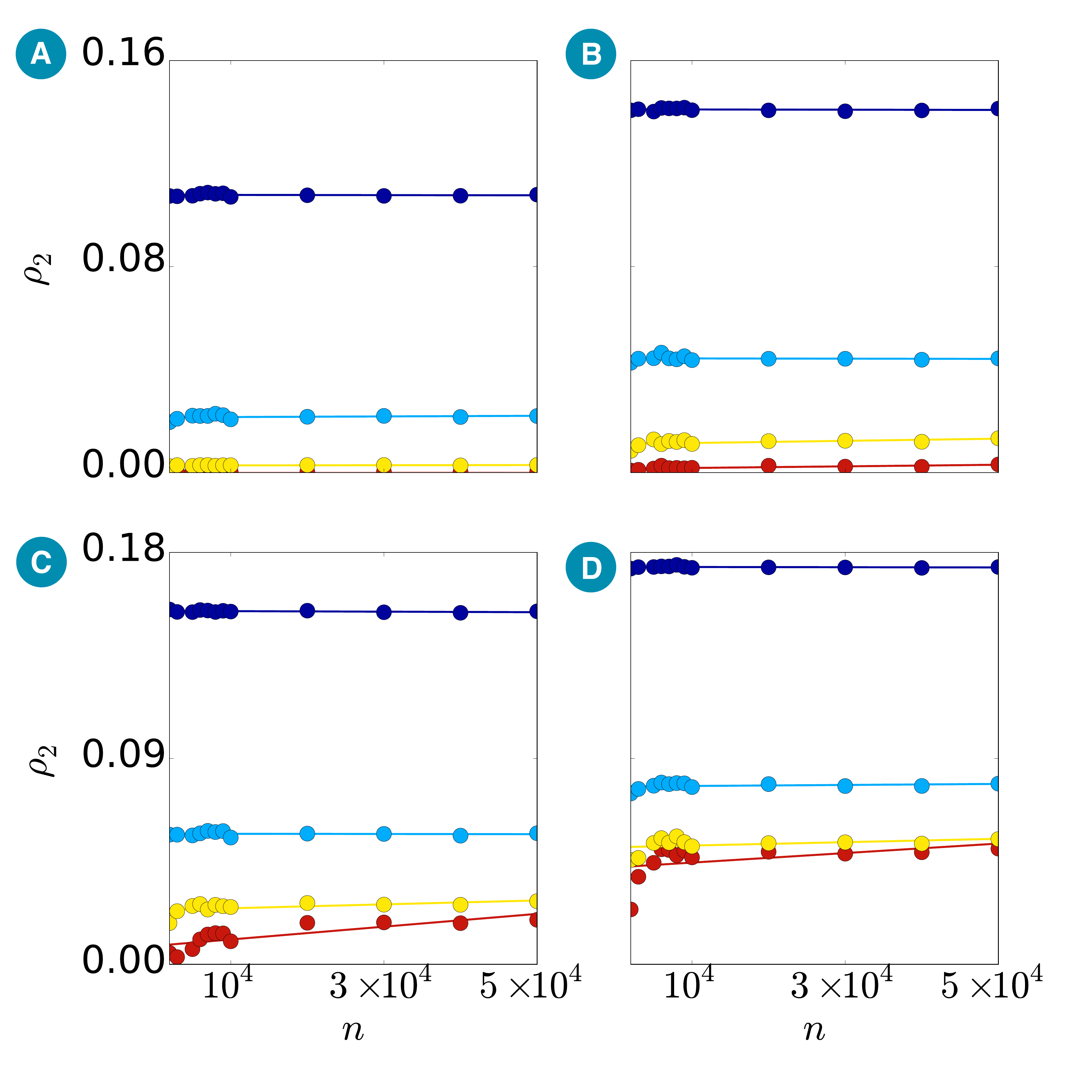}
\end{center}
\caption{Final fraction of infected nodes on the layer with lowest individual eigenvalue as a function of the the size of the layers. The colors represent different values of $\eta$, while on we have  $\lambda = 0.078$ on (a), $\lambda = 0.083$ on (b) $\lambda = 0.085$ on (c) and $\lambda = 0.088$ on (d). Furthermore, the layer sizes are $n = 2 \times 10^{3}, 3 \times 10^{3}, 4 \times 10^{3}, 5 \times 10^{3}, 6 \times 10^{3}, 7 \times 10^{3}, 8 \times 10^{3}, 9 \times 10^{3}, 10^{4}, 2 \times 10^{4}, 3 \times 10^{4}, 4 \times 10^{4}$ and $5 \times 10^{4}$ and $m = 2$ on all cases. Each curve is the result of a parameter $\eta$, from bottom to top $\eta = 10^{-4}, 10^{-3}, 10^{-2}, 10^{-1}$.}
\label{fig:FSS2}
\end{figure}

\subsubsection{Identical layers} \label{sec:app_identical}

Considering a multiplex network made up of two layers with the same configuration. Each layer of the multiplex is a network composed by $n = 1000$, $\langle k \rangle \approx 6$, $\Lambda^l = 14.34$, with degree distribution $P(k) \sim k^{-2.7}$. Aside from the intra-edge configuration, we also impose that inter-edges connect a node with its counterpart on the other layer, i.e., every node has the same intra-degree on all layers. Such a constraint imposes a high correlation between the degrees on each layer.

Figure~\ref{fig:Localization_Identical} shows the spectral behavior of such a multiplex as a function of the parameter $\left( \frac{\eta}{\lambda} \right)$. On the top panel, we represent the inverse participation ratio of the first three eigenvalues, while on the bottom panel, we plot the first ten eigenvalues. When the ratio $\frac{\eta}{\lambda} = 0$ the eigenvalues have multiplicity two, as can be seen on the left side of the bottom panel (approximately, since the figure starts from $10^{-2}$). More importantly, those eigenvalues tend to behave differently: one increases, while the other tends to decrease. This behavior leads to the eigenvalue crossing (see Appendix~\ref{sec:app_crossing}). The inset of the bottom panel zooms out the region where the crossing takes place. Note that the eigenvalues cross at the same value for which the inverse participation ratio shows an abrupt change. Indeed, the jump in the $\text{IPR}(\Lambda)$ has its roots in the interchange of the eigenvectors associated to each of the eigenvalues that are crossing. Moreover, we stress that the abrupt change observed for $\text{IPR}(\Lambda)$ is always present in such scenarios, but it could be either from the lower to the higher values or vice versa depending on the structure of the layers.

\subsubsection{Similar layers} \label{sec:app_similar}

In addition to the identical case, we have also considered a multiplex network composed by two layers with the same degree distribution (i.e. the same degree sequence), with $P(k) \sim k^{-2.7}$, but different random realizations of the configuration model. Furthermore, the inter-edges follow the same rule as before, connecting nodes with their counterparts on the other layer assuring that every node has the same intra-degree on all layers. Each layer of the multiplex network is composed by $n = 1000$ and $\langle k \rangle \approx 6$. Since each layer is a different realization of the configuration model, both present a slightly different leading eigenvalue, the first $\Lambda^1_1 = 15.21$ and the second $\Lambda^2_1 = 14.34$.

Figure~\ref{fig:Localization_Similar} shows the spectral behavior of such a multiplex in terms of the largest eigenvalues, on the bottom panel, and the $\text{IPR}(\Lambda)$, on the top panel. Here, in addition to the global inverse participation ratio, we also present the contribution of each layer to this measure. Such analysis is meaningless on the identical case, since the contribution is the same. As shown in the figure, we observe that for small values of $\frac{\eta}{\lambda}$, in regard to the first eigenvalue, the system is localized on the first layer and delocalized on the second. On the other hand, the picture changes when we focus on the second eigenvalue, as it is localized on the second layer, but delocalized on the first. For larger values of $\frac{\eta}{\lambda}$, both layers contribute equally to $\text{IPR}(\Lambda)$. Analogously to the identical case, there is a change on $\text{IPR}(\Lambda_2)$, which seems to be related to the changes on $\Lambda_2$, as one can see on the bottom panel and in the inset. Note that for this case, there is no crossing, i.e., the eigenvalues avoid the crossing -also referred to as near-crossing.

\begin{figure}[!t]
\begin{center}
\includegraphics[width=1\linewidth]{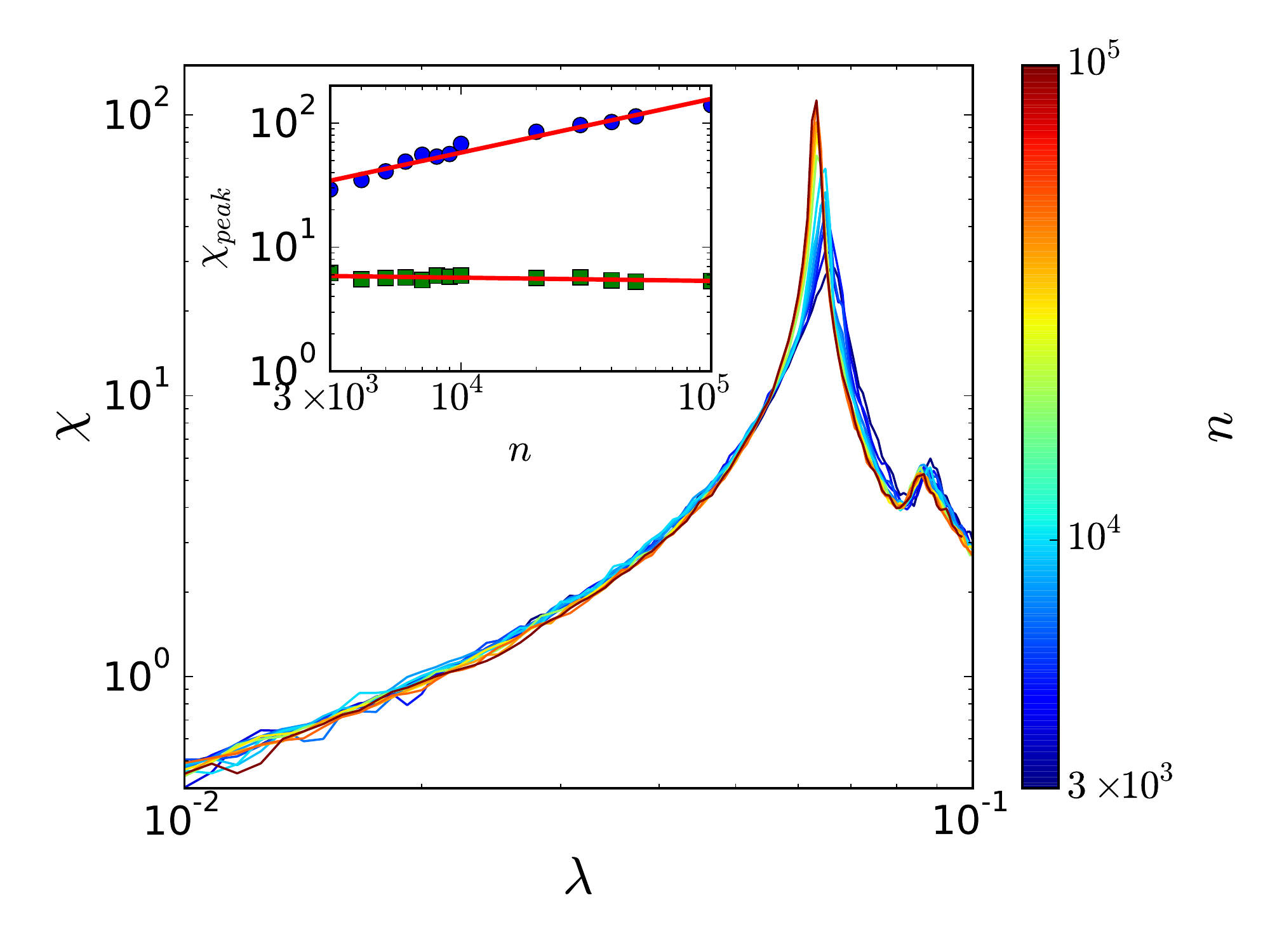}
\end{center}
\caption{Finite size analysis of the susceptibility. On the main panel we have the susceptibility as a function of $\lambda$ for different sizes of 2 layer multiplex network, where the first layer have $\langle k \rangle = 16$ and the second $\langle k \rangle = 12$. On this experiment we fixed the ratio $\frac{\eta}{\lambda} = 0.01$. On the inset we show the susceptibility of the two peaks as a function of the layer size, where the blue symbols refer to the first peaks, while the green symbols refer to the second peak. Besides, the red lines are a linear fitting of those points. The layer sizes evaluated are $n = 3 \times 10^3, 4 \times 10^3, 5 \times 10^3, 6 \times 10^3, 7 \times 10^3, 8 \times 10^3, 9 \times 10^3, 10^4, 2 \times 10^4, 3 \times 10^4, 4 \times 10^4, 5 \times 10^4, 10^5$.}
\label{fig:X_FSS}
\end{figure}

% Spectra 3 Layers
\begin{figure*}[!t]
\begin{center}
\includegraphics[width=1\linewidth]{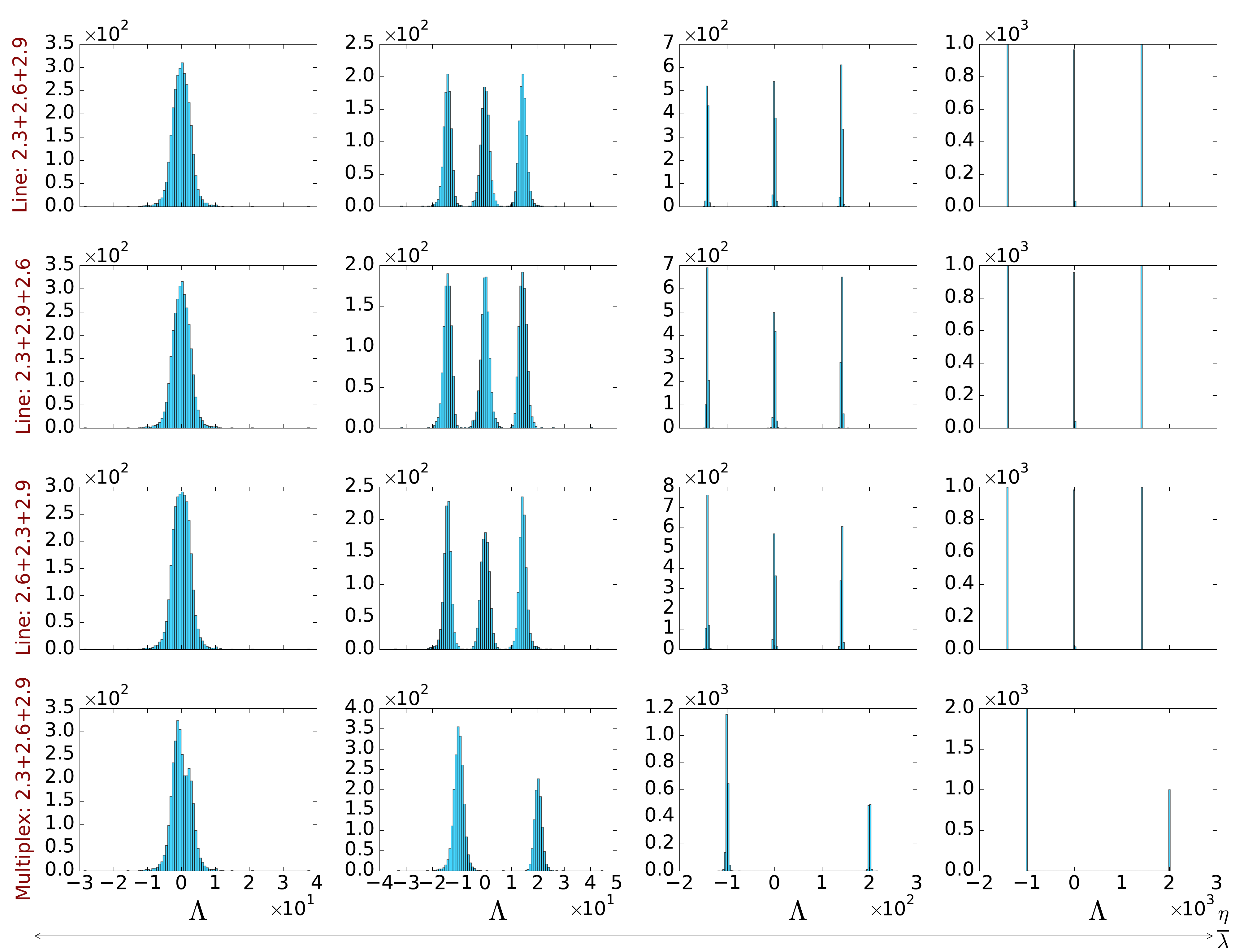}
\end{center}
\caption{Distribution of the eigenvalues. On the rows, from top to bottom, for the interconnected networks of Lines $2.3+2.6+2.9$, $2.3+2.9+2.6$, $2.6+2.3+2.9$ and the multiplex. On the columns, from left to right, we varied the ratios $\frac{\eta}{\lambda} = 1, 10, 100$ and $1000$ respectively. All histograms were built with 100 bins.}
\label{fig:Spectrum_3Layer}
\end{figure*}

\begin{figure}[!t]
\begin{center}
\includegraphics[width=1\linewidth]{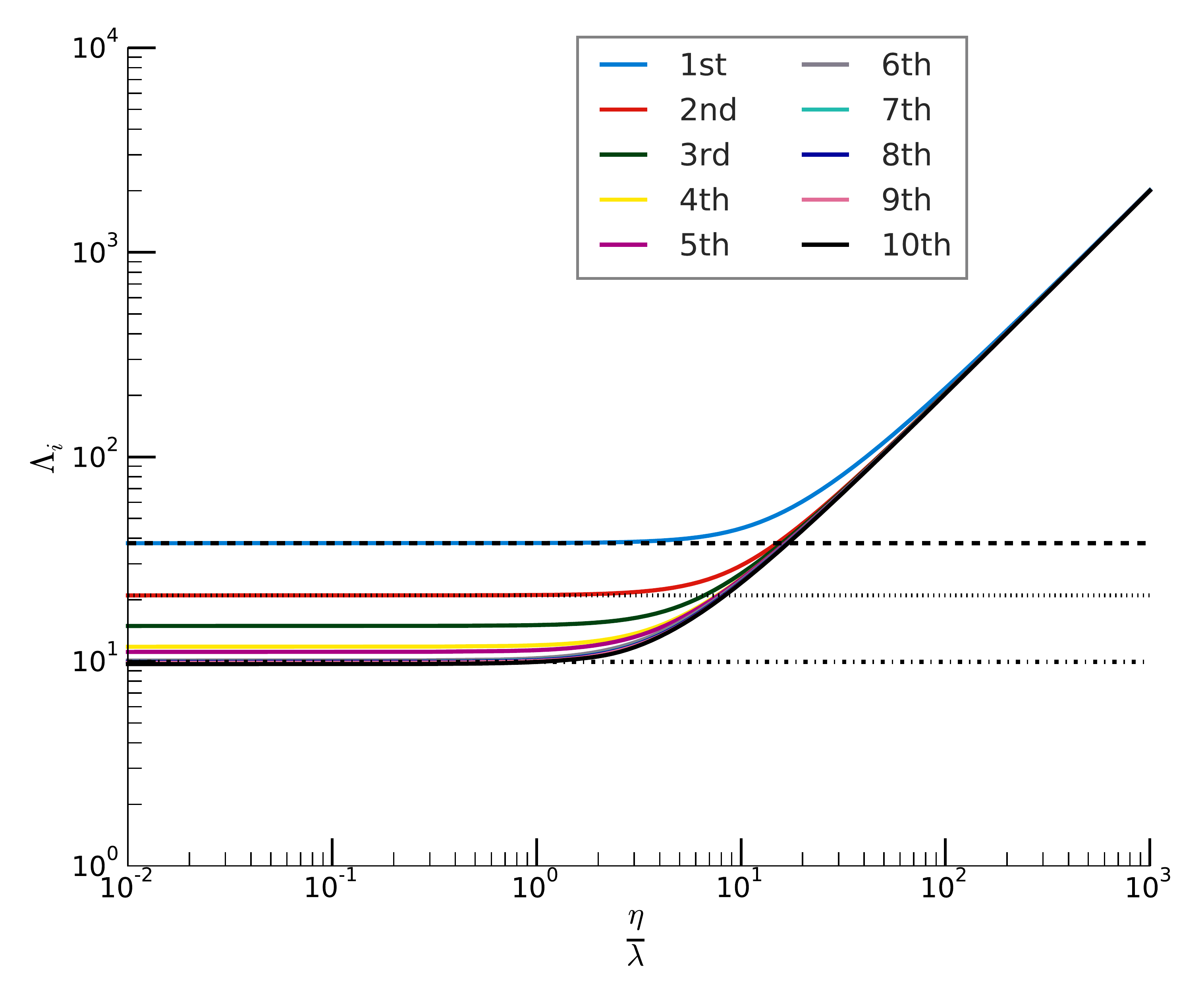}
\end{center}
\caption{Evaluation of the 8 first eigenvalues of $\mathcal{R}(\lambda, \eta)$ for the multiplex configuration as a function of of the ratio $\frac{\eta}{\lambda}$. It is noteworthy that such plot is visually equivalent for all the layer topologies composed by 3 layers. The dashed lines represents the individual layer leading eigenvalues.}
\label{fig:Eigs}
\end{figure}

\subsection{Finite size analysis} \label{sec:app_fss}

In this section we analyze the behavior of a 2-layer multiplex network at the steady state considering different sizes. Such a multiplex was built considering two Erd\H{o}s -- R\'enyi networks with a fixed mean degree. As mentioned in the main text, we chose this type of networks because their epidemic threshold do not vanish at the thermodynamic limit, which contrasts with the scale-free networks. In this way, we have a well-defined critical point that can be precisely tuned regardless of the network size. Following the usual convention on the complex network literature, the first susceptibility peak observed on our experiments can be classified as a critical point of a phase transition. On such point, the dynamics goes from a disease-free state to an endemic state. However, the second susceptibility peak cannot be classified as a second order phase transition, since the disease is already in an endemic state. Although it cannot be considered as a critical point, before the second susceptibility peak most of the events take place on only one layer (the one with the largest individual eigenvalue), while after this point both layers are active and spreading the disease.

Similarly to the experiments shown in Section~\ref{sec:sec_peak}, here  we run the continuous simulation 50 times and perform a moving average filter over a sampling of the original time series, resulting in $5\times10^4$ points. The simulations are run up to $t = 10^3$. Note that for continuous simulations the number of points can vary from one run to another. The steady state statistics are estimated for $t \geq 950$ or in other words, the last 50 time units. In contrast with the main text, here we are interested in comparing results for different network sizes, $n = 2 \times 10^{3}, 3 \times 10^{3}, 4 \times 10^{3}, 5 \times 10^{3}, 6 \times 10^{3}, 7 \times 10^{3}, 8 \times 10^{3}, 9 \times 10^{3}, 10^{4}, 2 \times 10^{4}, 3 \times 10^{4}, 4 \times 10^{4}$ and $5 \times 10^{4}$ and $m = 2$ in all cases. Besides, we considered the mean degree as $\langle k \rangle = 16$ for the first layer and $\langle k \rangle = 12$ for the second. We expect that the second susceptibility peak appears near the epidemic threshold of the second layer individually, i.e. $\lambda \approx 0.083$.

Figure~\ref{fig:FSS} presents the number of infected nodes in the steady-state on the layer with the lowest individual eigenvalue as a function of the size of the layers and a combination of the parameters $\lambda = 0.078, 0.083, 0.085, 0.088$ (near the individual critical point of the second layer) and $\eta = 10^{-4}, 10^{-3}, 10^{-2}, 10^{-1}$. Besides, on the insets we have the information about the average fraction (left inset on each panel) and its fluctuations, measured by the standard deviation (right inset on each panel). The straight lines in red were obtained by a least squares regression method.

We observe an approximately linear behavior of the number of infected nodes on the second layer as a function of the number of nodes on such layer (see the main panels of Fig.~\ref{fig:FSS}). Consequently, the fraction $\rho_2$ also presents a linear trend (see the left inset on each panel of Fig.~\ref{fig:FSS}). In fact, it presents a flat pattern, i.e approximately constant. Besides, the number of infected nodes is always larger than zero, since it is not a disease-free state. Furthermore, we also observed that the fluctuations tend to be very low (see the right inset on each panel of Fig.~\ref{fig:FSS}). Regarding the fluctuations, it is noteworthy that on a phase transition they tend to diverge, which does not happen in our analysis, thus also ruling out a second order phase transition as far as it concerns. We also note that fluctuations are slightly higher for lower spreading rates, as can be seen by the error bars for $\lambda = 0.078$, which is explained by the delocalization of our system.

Furthermore, in figure~\ref{fig:FSS2} we present the comparison of steady state fractions. In each panel, we fix a value of $\lambda$ and compare different values of $\eta$. It emphasizes the influence of $\eta$ on the final fraction of infected nodes on the second layer. Note that for $\eta = 10^{-4}$ and small networks the behavior exhibits a growing trend. This is due to the fact that for networks with $n < 10^4$ the contribution of the first layer can be effectively neglected. In fact, observe that for $n \geq 10^4$ the fraction of infected nodes on the second layer follows a flat pattern (see Fig.~\ref{fig:FSS2} (c) and (d)). Finally, in figure~\ref{fig:X_FSS} we present a finite size analysis of the susceptibility for different sizes, ranging from $n = 3 \times 10^3$ to $n = 10^5$ and $m = 2$ layers. Each curve was obtained using the QS algorithm, with which we simulated 120 points from $\lambda = 10^{-2}$ to $\lambda = 10^{-1}$. On such experiment we fixed the ratio $\frac{\eta}{\lambda} = 0.01$. Additionally, we also used a moving average filter with two points for visualization purposes. In the inset, we show the scaling of the susceptibility corresponding to the two peaks. The positive slope for the first peak indicates that it divergences as the system size goes to infinity, thus evidencing the phase transition. On the other hand, the curve for the case of the second peak is flat whatever the value of the system size is, indicating that in contrast to the behavior observed for the first peak, in this case there is no divergence in the thermodynamic limit nor the peak vanishes.

\section{3-Layer interconnected systems: complementary analysis} \label{sec:app_3L}

In this section we study the introduction of a third layer, which increases the complexity of the system allowing four  different network layer configurations, the line, which has three different configurations depending on the position of the layers, and the triangle, which is also a multiplex. This section is organized as follows: in the first subsection we perform the spectral analysis of the adjacency tensor as a function of the parameter $\frac{\eta}{\lambda}$, showing that as we increase this parameter the spectral distribution tends to the spectra of the network of layers, which is explained by interlacing theorems. Next, on sections~\ref{sec:app_loc_3L} and~\ref{sec:app_X_3L} we show the complementary results of localization and susceptibility analysis, respectively.

% Multiple peaks
\begin{figure}[!t]
\begin{center}
\includegraphics[width=1\linewidth]{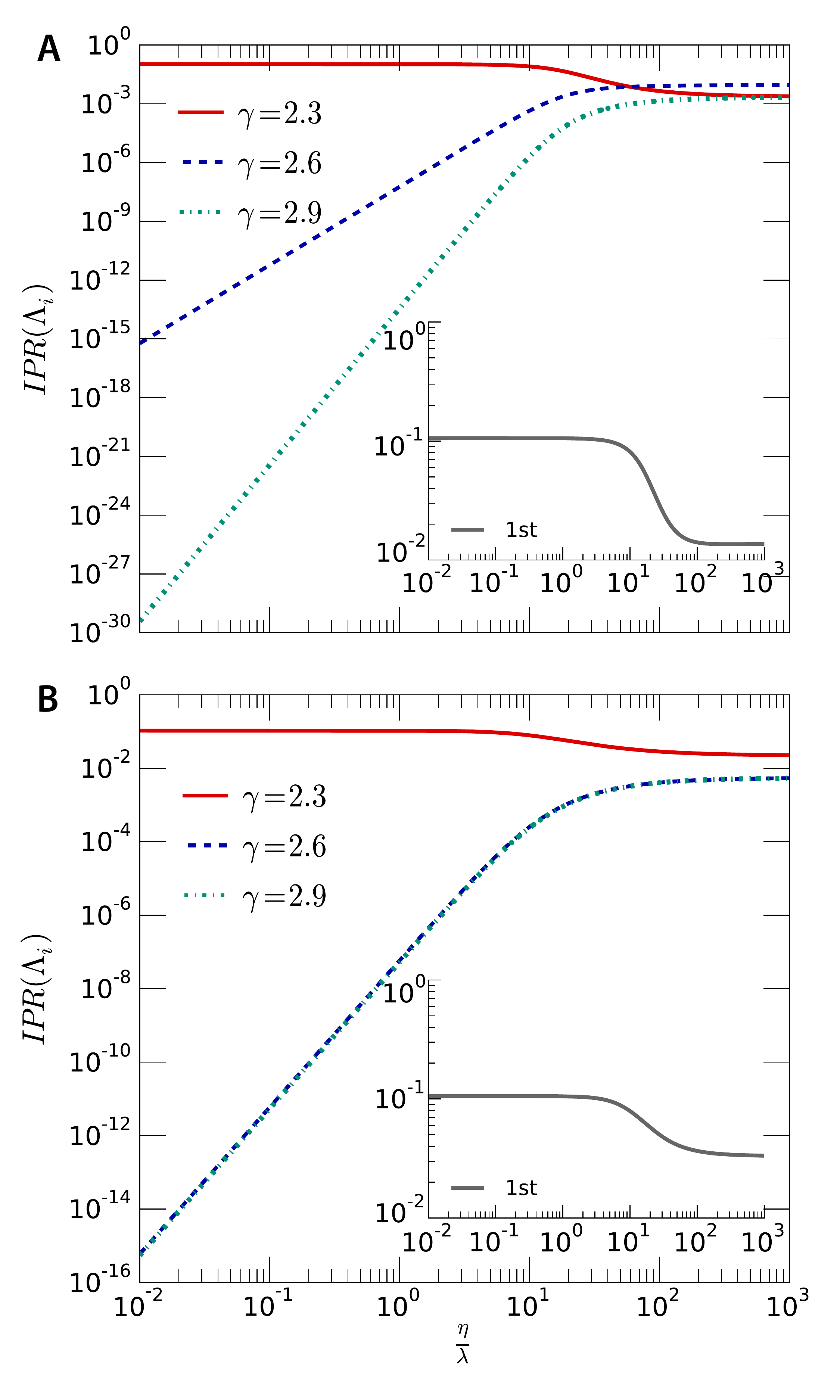}
\end{center}
\caption{Spectral properties of the tensor $\mathcal{R}(\lambda, \eta)$ as a function of the ratio $\frac{\eta}{\lambda}$ for a multiplex with two layers with the same degree distribution (different random realizations of the configuration model) and connected to its counterpart on the other layer. On the top panel we present the inverse participation ratio ($\text{IPR}(\Lambda)$) of the two larger eigenvalues and the individual layer contributions, while on bottom panel we show the leading eigenvalues. Every curve is composed by $10^3$ log spaced points, in order to have enough resolution.}
\label{fig:IPR_Line_23_26_29_SF_Line_26_23_29_SF}
\end{figure}

\begin{figure}[!t]
\begin{center}
\includegraphics[width=1\linewidth]{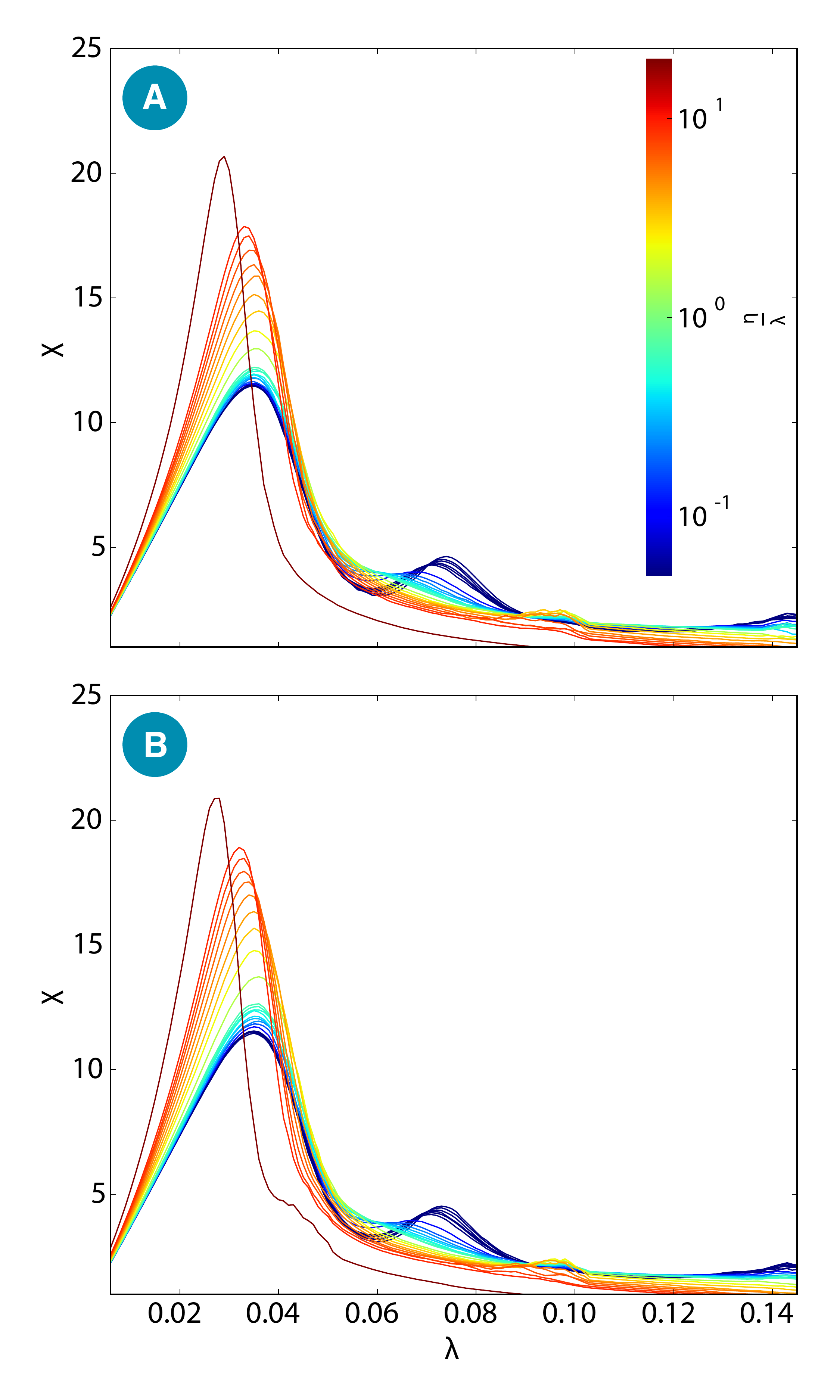}
\end{center}
\caption{Susceptibility $\chi$ as a function of $\lambda$ considering all three layer configurations and many different ratios $\frac{\eta}{\lambda}$, which is represented by the color of the lines. The recovering rate is $\mu=1$. The simulated values are $\frac{\eta}{\lambda} = $ 0.05, 0.06, 0.07, 0.08, 0.09, 0.1, 0.2, 0.3, 0.4, 0.5, 0.6, 0.7, 0.8, 0.9, 1.0, 2, 3, 4, 5, 6, 7, 8, 9, 10, 20.}
\label{fig:X_Line_23_26_29_SF_Line_26_23_29_SF}
\end{figure}

\subsection{Spectral analysis} \label{sec:app_3L_spec}

Since the epidemic process is described through the supra adjacency tensor $\mathcal{R}(\lambda, \eta)$, its spectral properties give us some insights about the whole process, especially about the critical properties of the systems under analysis. Moreover, as the structure of the network of layers is not trivial anymore, we shall find important differences regarding the spectra of such tensors for the different topologies of the network of layers.

Figure~\ref{fig:Spectrum_3Layer} shows the spectrum of the four configurations of networks when varying the ratio $\frac{\eta}{\lambda} = 1, 10, 100$ and $1000$. Observe that we do not show the ratio $\frac{\eta}{\lambda} = 0$ since it is just the union of the individual layers' spectrum. For $\frac{\eta}{\lambda} = 1$, the four configurations are very similar, especially the line graphs. In such case, the inter-layer edges are treated in the same way as the intra-layer ones. In other words, they are ignored and the network can be interpreted as a monoplex network. As the spreading ratio increases the spectrum tends to be clustered near the values of the eigenvalues of the network of layers. Such spectra was analytically calculated in Section~\ref{sec:teo} and shown in Table~\ref{tab:spec_net_net} on the main text.

Regarding the triangle configuration, the clustering of the spectrum as
$\frac{\eta}{\lambda}$ increases is even clear. Triangles present the
lowest eigenvalue with multiplicity two. On the extreme case of
$\frac{\eta}{\lambda} \gg 1$, see Fig.~\ref{fig:Spectrum_3Layer}, we
have $2/3$ of the values near the left extreme value while $1/3$ is near
the leading eigenvalue. On the other hand, for the line configurations,
the frequencies of the eigenvalues distribution is related to the
position of the central layer. However, on the limiting cases such
differences are reduced. This pattern is naturally related
to the increase of the spreading ratio: When $\frac{\eta}{\lambda}$
increases, so does the role of the inter-layer edges relative to the
intra-layer ones. Consequently, the structure of the network of layers
imposes itself more strongly on the eigenvalues of the entire
interconnected structure. This comes as a consequence of the interlacing
theorems shown in Section~\ref{sec:Spectra_interlacing} on the main text.

Our findings can be related to the structural transition shown in~\cite{Radicchi2013}, where the authors evaluated the supra-Laplacian matrix as a function of the inter-layer weights. Their main result is an abrupt structural transition from a decoupled regime, where the layers seem to be independent, to a coupled regime where the layers behave as one single system. Here, we are interested in the supra-adjacency tensor, however, we found a similar phenomenological behavior and a structural change of the system as a function of the inter-layer weights, which in our case are determined by a dynamical process.

\subsection{Localization on interconnected networks} \label{sec:app_loc_3L}

Complementary to the results presented in Section~\ref{sec:3L}, here we present results for the lines $(2.3+2.6+2.6)$ and $(2.6+2.3+2.6)$. Similarly, the experiments here are conducted in terms of the inverse participation ratio, as it was done for the 2-Layer multiplex case. 

Figure~\ref{fig:Eigs} shows the 10th larger eigenvalues of the 3-layer multiplex case. The dashed lines represent the leading eigenvalue of each layer. Note that the leading eigenvalue of the layer with $P(k) \sim k^{-2.9}$ is the 7th larger on the network spectrum when $\frac{\eta}{\lambda} = 0$. We observe that there is no crossings on the observed eigenvalues, which is an expected result, since the layers have different structures. Furthermore, it is important to remark that all networks of layers evaluated also show similar qualitative behaviors. The topology of the network of layers does not lead to qualitative differences on the dependence of $\Lambda_i$ on $\frac{\eta}{\lambda}$ for the first ten eigenvalues. We also notice that although it is only an approximation, the perturbation theory would be valid roughly up to $\frac{\eta}{\lambda} \lesssim 10$. 

Figures~\ref{fig:IPR_Triangle_Line_23_29_26_SF} and~\ref{fig:IPR_Line_23_26_29_SF_Line_26_23_29_SF} shows the $\text{IPR}(\Lambda_1)$. On the main panel we present the individual contribution of each layer, while on the insets we have the total $\text{IPR}(\Lambda_1)$. As mentioned on the main text, the first eigenvalue is usually enough to analyze the localization as a first order approximation. Here we observe that the layer with the largest eigenvalue dominates the dynamics. In addition, note the similarities between the multiplex and the line configuration $(2.6+2.3+2.6)$, where the non-dominant layers behave similarly. This is because for small values of $\frac{\eta}{\lambda}$, the effect of the extra edge in the network of layers (closing the triangle) is of order $\eta^2$ and so the similar behavior observed comparing the panel (b) of figures~\ref{fig:IPR_Triangle_Line_23_29_26_SF} and~\ref{fig:IPR_Line_23_26_29_SF_Line_26_23_29_SF} for the two configurations. As $\frac{\eta}{\lambda}$ grows, the symmetry in the node-aligned multiplex dominates the eigenvector structure and the contributions of all layers are comparable. As we next show, the different contributions of the layers to the total $\text{IPR}(\Lambda_1)$ are at the root of the multiple susceptibility peaks observed.

Complementing and reinforcing the analysis of Section~\ref{sec:3L}, comparing the different line configurations of the network of layers, observe that the largest eigenvalue of the whole system, $\Lambda_1$, has its associated eigenvector localized in the dominant layer, that is, in the layer generated using $\gamma=2.3$. Depending on the position of that layer in the whole system --- i.e., central or peripheral layer ---, the contribution of the non-dominant layers to $\text{IPR}(\Lambda_1)$ varies. In particular, when the dominant layer corresponds to an extreme node of the network of layers, the contribution of the other two layers will ordered according to the distance to the dominant one. Consequently, when the dominant layer is in the center of the network of layers, the contributions of the non-dominant ones are comparable -note that in panel (b) of Fig.~\ref{fig:IPR_Line_23_26_29_SF_Line_26_23_29_SF}, there is no difference in the contribution to $\text{IPR}(\Lambda_1)$ of layers generated using $\gamma=2.6$ and $\gamma=2.9$.

\subsection{Multiple susceptibility peaks: additional results} \label{sec:app_X_3L}

Figure~\ref{fig:X_Line_23_26_29_SF_Line_26_23_29_SF} shows the susceptibility as a function of $\lambda$ for different ratios of $\frac{\eta}{\lambda}$. As observed in the main text, we also have three well-defined peaks in these curves when the ratio $\frac{\eta}{\lambda}$ is small. In addition, similar to the 2-layer case, such peaks tend to become less defined and vanish as the ratio $\frac{\eta}{\lambda}$ increases. 

Regarding the third peak, note that it is less defined than the others because the average number of infected nodes is larger in this case. Consequently the susceptibility tends to be lower, since it measures the variance in relation to the average. The comparison of Figures~\ref{fig:X_Triangle_Line_23_29_26_SF} (b) and~\ref{fig:X_Line_23_26_29_SF_Line_26_23_29_SF} shows that there is no difference in the position of the susceptibility peaks. As mentioned in the main text, the only observed difference is the barrier effect, shown in Fig.~\ref{fig:X_Triangle_Line_23_29_26_SF} (a). We also remark the similarities between the line $(2.6+2.3+2.6)$ and the multiplex case, which emphasize the role of the central node. In that line configuration, the layer with $\gamma = 2.3$ spreads its influence to both layers, being this similar to the multiplex case, however with less intra-edges.

\bibliography{paper}

\end{document}